\documentclass[prd,showpacs, onecolumn, nofootinbib, preprintnumbers]{revtex4}%

\usepackage{amssymb,amsmath}
\usepackage[dvips]{color}
\usepackage[dvips]{graphicx}
\usepackage{epsfig}

\begin{document}

\preprint{JLAB-THY-07-689}

\title{Charmonium excited state spectrum in lattice QCD}

\author{Jozef J. Dudek}
\affiliation{Jefferson Laboratory MS 12H2, 12000 Jefferson Avenue, Newport News, VA 23606, USA}
\affiliation{Department of Physics, Old Dominion University, Norfolk, VA 23529, USA}
\email{dudek@jlab.org}

\author{Robert G. Edwards}
\affiliation{Jefferson Laboratory MS 12H2, 12000 Jefferson Avenue, Newport News, VA 23606, USA}
\email{edwards@jlab.org}

\author{Nilmani Mathur}
\affiliation{Jefferson Laboratory MS 12H2, 12000 Jefferson Avenue, Newport News, VA 23606, USA}
\email{dgr@jlab.org}

\author{David G. Richards}
\affiliation{Jefferson Laboratory MS 12H2, 12000 Jefferson Avenue, Newport News, VA 23606, USA}
\email{dgr@jlab.org}

\begin{abstract}

Working with a large basis of covariant derivative-based meson interpolating
fields we demonstrate the feasibility of reliably extracting multiple excited states
using a variational method. The study is performed on quenched
anisotropic lattices with clover quarks at the charm mass. We
demonstrate how a knowledge of the continuum limit of a lattice
interpolating field can give additional spin-assignment
information, even at a single lattice spacing, via the overlap factors
of interpolating field and state. Excited state masses are
systematically high with respect to quark potential model predictions
and, where they exist, experimental states. We conclude that this is
most likely a result of the quenched approximation.

\end{abstract}


\maketitle 

\section{Introduction}\label{intro}

In this paper we discuss an application of the variational method for
extraction of spectroscopic information from lattice QCD two-point
functions. We focus on the use of a large set of interpolating fields
allowing access to all $J^{PC}$ with sufficient redundancy to consider
several excited states in each channel. As well as the usual
extraction of mass information, we are also interested in the overlaps
of interpolating fields on to states in the spectrum. This information
is required if one wishes to extract matrix elements between hadron
states from three-point or higher correlators, an example being the
radiative transitions considered in \cite{Dudek:2006ej}. We find in this
paper that these overlaps are also useful in aiding the
spin-identification of states provided one has established the
continuum behavior of the interpolating fields.

Our application of the variational method utilizes the orthogonality
of states to extract excited state information. This is particularly
powerful when a number of excited states are close to being degenerate
- conventional multi-exponential fitting methods are not good at
resolving such states.  We are able to collect high statistics by the
use of a small spatial volume $\sim (1.2 \mathrm{fm})^3$ and by
confining our study to a large pseudoscalar mass $m_{\rm{ps}} \sim 3
\,\rm{GeV}$, corresponding to charmonium. This volume is known to be
sufficient to house the ground state for most $J^{PC}$, as illustrated
for the ``charge'' radii extracted in \cite{Dudek:2006ej}, but we might
expect it to be insufficient for radially excited states and those of
high spin. We will study the volume dependence of a number of
correlators and find, somewhat surprisingly, that there seems to be no
meaningful difference between a $1.2\,\rm{fm}$ box and a
$2.4\,\rm{fm}$ box.

A resurgence of interest in charmonium physics is underway, spurred by
new results from BaBar, Belle and CLEO. This is reflected in increased
theoretical effort, in particular there have been recent efforts to
apply lattice QCD techniques to computation of quantities previously
the reserve of potential models and sum-rules, such as radiative
transitions\cite{Dudek:2006ej} and two-photon
decays\cite{Dudek:2006ut}. In this paper we extract information about
the excited state and high spin spectrum, which is both interesting in
its own right, and vital for concurrent attempts to extract excited state
radiative transitions.

We begin in section II with a discussion of the lattice action and other
computational details before in section III outlining our choice of a broad set of
interpolating fields constructed to be irreducible representations of
the lattice symmetry group at zero momentum and to have simple, known
continuum properties. The variational method used to describe the data
is explained in section IV with focus on ensuring that the solution accurately
describes the input data. Spectrum results are then shown in section V, with some
discussion of how continuum spin assignments can be aided by
considering overlap factors. In three appendices we display properties
of our interpolating fields and show an application of the variational
method to toy data.

\section{Computational Details}\label{sim}

The computations were performed in the quenched approximation to QCD,
using the Chroma software system~\cite{Edwards:2004sx}.
We employed 1000 configurations on a $12^3 \times 48$ lattice generated
using an anisotropic Wilson gauge action~\cite{Klassen:1998ua}
\begin{equation}
  S  ~=~ -\beta \Bigl(\, \frac{1}{\xi_0} \sum_{x,i>j} P_{ij}(x) \,+\,
                         \xi_0  \sum_{x,i} P_{0i}(x)\Bigr)\quad,
\label{anisoW}
\end{equation} 
defined in terms of simple plaquettes $P_{\mu\nu}(x)$. Here, $\beta$
is the bare coupling, the time component is labeled by $0$, the
spatial indices $i,j$ run from $1$ to $3$, and $\xi_0$ is the bare
anisotropy. We tune $\xi_0$ to the desired renormalized anisotropy
$\xi \equiv a_s/a_t = 3$ where $a_s$ and $a_t$ are the spatial and
temporal lattice spacings, respectively and find $\xi_0 = 2.464$.  The temporal lattice spacing
obtained from the static quark-antiquark potential is $a_t^{-1} = 6.05
(1)~{\rm GeV}$. This gauge action is expected to have
${\cal{O}}(a_s^2,a_t^2)$ discretization errors. A smaller set of 807 configurations on a $24^3\times 48$ lattice were used to study
finite volume effects.

Anisotropic lattices as applied to charmonium exploit the fact that
while the quark mass scale demands a cut-off above $\sim 1.5~
\mathrm{GeV}$, the internal three-momentum scale is typically much lower,
$\sim 500~\mathrm{MeV}$. On our lattice, we can have both $m_c a_t$
and $| \vec{p} | a_s$ reasonably small and a spatial
length $\gtrsim 1~\mathrm{fm}$ without requiring very many spatial
lattice sites. 
Most of the work presented uses one volume, $L_s \approx 1.2~
\mathrm{fm}$ while a limited comparison is made with a large volume
$L_s\approx 2.4~\mathrm{fm}$; previous charmonium spectrum studies indicate that
there are no significant finite volume effects for lattices of this
size or larger\cite{Choe:2003wx,Drummond:1999db}.

The quark propagators were computed using an anisotropic version of
the ${\cal{O}}(a)$ tadpole-improved Clover
action~\cite{Sheikholeslami:1985ij,Klassen:1998fh,Chen:2000ej}.  The
Clover action we used has the ``mass'' form with spatial fermion
tuning. Defining a lattice spacing $a_\mu$ where $a_k=a_s$ and
$a_0=a_t$, we can express this fermion action in terms of dimensionless
variables $\hat \psi = a_s^{3/2} \psi$, $\hat W_\mu = a_\mu W_\mu$ and
$\hat F_{\mu\nu} = a_\mu a_\nu F_{\mu\nu}$ as
\begin{eqnarray}
a_t Q = a_t m_0 + \hat W_0 \gamma_0 +
 \frac{\nu}{\xi_0} \sum_k \hat W_k \gamma_k -
 \frac{1}{2} \bigg[ c_t \sum_k \sigma_{0k} \hat F_{0k} +
 \frac{c_s}{\xi_0} \sum_{k<l} \sigma_{kl} \hat F_{kl} \bigg ]\quad.
\label{eq:Q_Tim}
\end{eqnarray}
The factor $\nu$ is the bare value of the fermion anisotropy.
For the field strength tensor
$F_{\mu\nu}$, we adopt the standard ${\cal{O}}(a_s^2,a_t^2)$ clover leaf
definition. Here, the ``Wilson'' operator has the
projector property
\begin{eqnarray}
W_\mu \equiv \nabla_\mu - \frac{a_\mu}{2} \gamma_\mu \Delta_\mu\quad,
\end{eqnarray}
where
\begin{eqnarray}
\nabla_\mu f(x) &=& \frac{1}{2a_\mu} \bigg[ U_\mu(x) f(x+\hat{\mu} a_\mu) -
 U^\dagger_\mu(x-\hat{\mu} a_\mu) f(x-\hat{\mu} a_\mu) \bigg ] \\
\Delta_\mu f(x) &=& \frac{1}{a_\mu^2} \bigg[ U_\mu(x) f(x+\hat{\mu}a_\mu) +
 U^\dagger_\mu(x-\hat{\mu}a_\mu) f(x-\hat{\mu}a_\mu) - 2f(x) \bigg ] .
\end{eqnarray}
The projector property ensures that no doubler states are present 
in the determined mass spectrum. We have tuned $(m_0, \nu)$ so as
to yield the desired quark mass and such that the speed of light
obtained from the meson dispersion relations be one, as discussed 
later. 

We have used the tree-level conditions on $c_s$ and $c_t$
as described in Ref.~\cite{Chen:2000ej}. In particular, we will choose
\begin{eqnarray}
c_s = \frac{\nu}{u_s^3}, \quad 
c_t = \frac{1}{2}\left(\nu + \frac{a_t}{a_s}\right)\frac{1}{u_t u_s^2}
\label{eq:chen_clover}
\end{eqnarray}
where the ratio $a_t/a_s = 1/\xi$ is set to the desired {\em
renormalized} gauge anisotropy. The tadpole factors $u_s$ and $u_t$
come from the fourth-root of the spatial and temporal plaquettes,
respectively, and take the values $0.814$ and $0.980$.

The charm mass is determined by tuning the bare quark mass $m_0$
non-perturbatively such that the spin average of the lowest $S$-wave
mesons coincides with its experimental value, i.e. 
$
(3 m_{J/\Psi} +
 m_{\eta_c})/4 = 3.067~{\rm GeV}.
$, such that $m_0 = 0.0401$.
We tune $\nu$ non-perturbatively to satisfy the
lattice dispersion relation
\begin{eqnarray}
 c(p)^{2} = {{E(p)^{2} - E(0)^{2}} \over p^{2}} = \xi^{2} {{a_{t}^{2} E(p)^{2} - a_{t}^{2}E(0)^{2}}\over a_{s}^{2}p^{2}} = 1.
\end{eqnarray}
Keeping all other parameters fixed, we tune $\nu$ to satisfy the
above relation to within $\sim 1\%$ and find $\nu = 0.867$.

We have used Dirichlet boundary conditions in the temporal direction
placing the source five time-slices from the wall. All subsequent
time-slice plots are with respect to this source position.

For comparison purposes, we have also used an anisotropic version of the
domain-wall fermion (DWF) action~\cite{Shamir:1993zy,Dudek:2006ej} with a
domain-wall height $a_t M = 1.7$, a fifth dimensional extent $L_5 = 16$,
and a quark mass $a_t m_q$.
The kernel of the domain action is the same as in Eq.~\ref{eq:Q_Tim} except
without the clover terms, i.e., $c_t$ and $c_s$ set to $0$.
As in the clover case, the domain-wall quark mass and fermion
anisotropy $(m_q, \nu)$ are chosen to yield the desired S-wave mass and
such that the speed of light obtained from the meson dispersion
relations is one.

\begin{figure}[h]
  \vspace{1cm}
       \psfig{width=10cm,file= 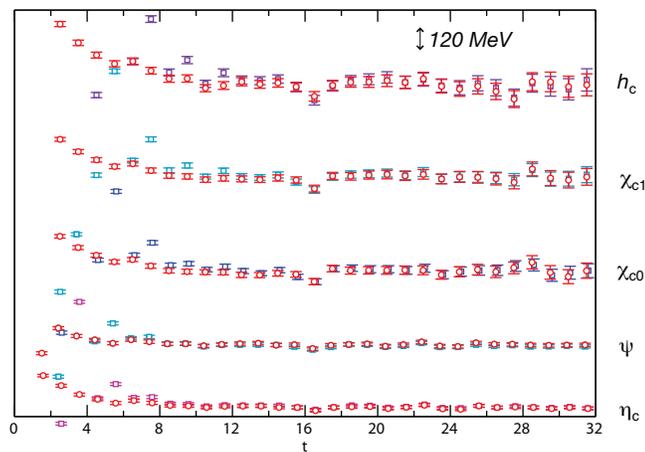}
      \caption{Effective masses for tadpole clover (red) and domain
        wall fermions (shades of blue/purple) on same lattices. Note
        masses are shifted for clarity. Relative mass scale is shown. }
  \label{fig:compare_dwf_cl}
\end{figure}

We have computed smeared-local correlators on a subset of the same quenched
lattices using both tadpole Clover and domain wall fermion quark
actions. The effective masses are shown in figure \ref{fig:compare_dwf_cl} where we see
that apart from the domain wall fermion oscillations at small times
there is essentially perfect agreement between the two actions. Since
the domain wall action should be ${\cal O}(a)$ improved automatically
we infer that, at least for the ground state spectrum, the tadpole
Clover is also ${\cal O}(a)$ improved to a good approximation.

In this study we used quenched lattices. In the spectrum of heavy
quark states, where unitarity violation due to the absence of closed charm quark
loops can be neglected, there are
likely to be two main effects of this approximation. Firstly there are
no virtual $D$ mesons in this theory and hence no thresholds for
OZI-allowed decay and secondly there is the known issue of the
incorrect running of the coupling and an associated ambiguity in the
setting of the lattice scale. In all cases we compute only the connected contributions to
correlators - the effects of disconnected contributions in charmonium
are believed to be small\cite{McNeile:2004wu, deForcrand:2004ia}.

\section{Interpolating Fields}

It is common in meson spectroscopy calculations to focus on the
simplest meson interpolating fields, namely the local fermion
bilinears, $\bar{\psi}(x) \Gamma \psi(x)$. By appropriately smearing
the operator over space, a large overlap on to the ground state of a given
quantum number can be obtained. These operators are, however, limited
in the $J^{PC} = 0^{\pm+}, 1^{+\pm}, 1^{--}$ with which they have
overlap. In addition, with only these operators, while one might try various different smearings
\cite{Burch:2006cc}, it is not possible to explore higher spin states. For these
reasons we opt to extend the operator basis by including non-local
operators, in particular ones constructed from a number of
covariant derivatives acting on the quark fields. On a discretised
lattice covariant derivatives become finite displacements of quark
fields connected by links.

In the continuum one can construct operators which transform in a
particular way under the rotation group, giving overlap only on to
certain spins. On a discretized lattice the full rotation group is
broken down to a smaller group of cubic rotations with a limited
number of irreducible representations (irreps). The operators used in this
paper are constructed to be both irreps of the lattice rotation group
(at zero momentum) and to have definite, simple forms in the continuum
limit. In this way we have some information about how these operators
will behave in the continuum limit - we will see that this potentially
aides in the interpretation of the lattice data if we are close
enough to restoring rotational symmetry.

Our operators are based upon those in \cite{Liao:2002rj}. The principal
differences are that we extend the basis to include all possible zero,
one and two derivative operators\footnote{although not all of these
  are actually used in this calculation}, generalize the derivatives so
that the operators have definite charge-conjugation at finite momentum
and correct a projection operator to ensure orthogonality within a
lattice irrep. Finally we develop the continuum overlap formalism
expressed in appendix \ref{overlaps}.

\subsection{Derivative-based operators}
In \cite{Liao:2002rj} single derivative operators of the type $\bar{\psi}_2
\Gamma_i \overrightarrow{D}_j \psi_1$ were utilized where 1 and 2
signify the flavor. The
right-facing arrow indicates that the covariant derivative acts only
on $\psi_1$. For the case of a single flavor we
would like the operators to be of definite charge-conjugation - it is easy to see using an integration by parts that
the above operators do not achieve this at finite momentum:
\begin{equation}
\int d^3 \vec{x} e^{i\vec{p}\cdot\vec{x}} \bar{\psi}(x) \Gamma_i
\overrightarrow{D}_j \psi(x) =
- \int d^3 \vec{x} e^{i\vec{p}\cdot\vec{x}} \bar{\psi}(x) \Gamma_i
\overleftarrow{D}_j \psi(x) + i p_j \int d^3 \vec{x}
e^{i\vec{p}\cdot\vec{x}} \bar{\psi}(x) \Gamma_i \psi(x).
\end{equation}
The first term on the right-hand side has the same charge-conjugation
as the left-hand side but the second term, lacking the derivative, has
the opposite $C$. This can be fixed by replacing the derivative
$\overrightarrow{D}$ by the quark-antiquark symmetrised 
$\overleftrightarrow{D} \equiv \overleftarrow{D} -
\overrightarrow{D}$. The same integration by parts as above will now
not yield a term proportional to the three-momentum.

At the two-derivative level, Ref. \cite{Liao:2002rj} construct two combinations,
$\mathbb{B}_i = \epsilon_{ijk} \overrightarrow{D}_j
\overrightarrow{D}_k$ and $\mathbb{D}_i = |\epsilon_{ijk}| \overrightarrow{D}_j
\overrightarrow{D}_k$. In the $\mathbb{D}$-type case we ensure definite
charge-conjugation at non-zero momentum by redefining $\mathbb{D}_i = |\epsilon_{ijk}| \overleftrightarrow{D}_j
\overleftrightarrow{D}_k$. The $\mathbb{B}$-type case does not actually require
this extension as one can easily see by expressing the pair of
derivatives as the sum of a commutator and an anticommutator:
\begin{equation}
 \epsilon_{ijk} \overrightarrow{D}_j \overrightarrow{D}_k =
 \epsilon_{ijk} \tfrac{1}{2}\Big(  [  \overrightarrow{D}_j ,
 \overrightarrow{D}_k ] + \{ \overrightarrow{D}_j ,
 \overrightarrow{D}_k \}  \Big) = \epsilon_{ijk} \tfrac{1}{2} [  \overrightarrow{D}_j ,
 \overrightarrow{D}_k ] = -\tfrac{i}{2} \epsilon_{ijk} F^{jk}.
\end{equation}
Clearly the $\mathbb{B}$-type operator is so named because it corresponds to the
chromomagnetic component of the field-strength tensor. We complete the two-derivative set by adding $\mathbb{E}_i = \mathbb{Q}_{ijk} \overleftrightarrow{D}_j
\overleftrightarrow{D}_k$ ($\mathbb{Q}_{ijk}$ is a Clebsch-Gordan
coefficient as defined in the next
section) and the Laplacian $\nabla^2 = \sum_i \overleftrightarrow{D}_i
\overleftrightarrow{D}_i $.

In the continuum, Lorentz symmetry along with parity and
charge-conjugation specify the form of the overlap of an operator with
a state of given $J^{PC}$. In appendix \ref{overlaps} we tabulate the forms, valid
for all three-momenta in Minkowski space-time, for the operators we use in this paper. 

We approximate covariant derivatives using finite displacements,
including the appropriate links. In the results presented below we displaced only by one
site, displacing by two sites was found to give correlators that did
not differ considerably. Our implementation takes the form:
\begin{equation}
\overrightarrow{\nabla}_j f(x) = \tfrac{1}{2a_s} \Big( U_j(x) f(x + \hat{j} a_s) -
U_j^\dag(x - \hat{j} a_s) f(x - \hat{j} a_s) \Big) \to \overrightarrow{D}_j f(x) + {\cal O}(a_s^2).
\end{equation}

\subsection{Projection onto lattice irreducible representations}
Our method will be much aided by having a basis of operators that transforms
irreducibly under the cubic group $O$. The irreducible representations of O, together with their continuum spin contents, are shown in Table~\ref{tab:irrep}~\cite{Johnson:1982yq}. The operators are constructed as
products of gamma matrices and derivatives as outlined above, and
therefore we need to project these products to their irreducible
components.  For operators $K$ and $L$ transforming according to the
irreps $\Lambda_K$ and $\Lambda_L$ respectively, the product $M$
transforms under an element $R$ of the cubic group as
\begin{equation}
M_{ij} \equiv K_i L_J \rightarrow \Gamma^{\Lambda_K}(R)_{i i'}
\Gamma^{\Lambda_L}(R)_{j j'} K_{i'} L_{j'}  \, ,\label{eq:product}
\end{equation}
where $\Gamma^{\Lambda}(R)$ is the representation matrix for the element
$R$ in the irrep $\Lambda$ of dimension $d_{\Lambda}$.  The essential
tool in constructing the irreps is the projection formula
\begin{equation}
{\cal O}_i^{\Lambda_{\alpha\beta}} = \frac {d_\Lambda}{g_{\cal O}} \sum_{R \in O}
\Gamma_{\alpha\beta}^{\Lambda}(R) U(R)_{ij} {\cal O}_j \, ,\label{eq:proj}
\end{equation}
where $\{{\cal O}_j\}$ is a basis of operators that is reducible under
the cubic group, $U(R)$ is the representation of the rotation $R$ on
that basis, and $\{{\cal O}^{\Lambda}_i \}$ are a set of operators
transforming irreducibly under the cubic group.  The indices
$\alpha$ and $\beta$ refer to the rows of the irreducible
representation, and the final step in the procedure is to identify a
set of linearly independent operators $\{{\cal O}^{\Lambda}_i: i =
1,\dots, d_{\Lambda}\}$ as a basis for the irreducible representation.

\begin{table}
\begin{tabular}{ccc}
$\Lambda$ & $d_{\Lambda}$ & $J$\\ \hline
$A_1$ & 1 & $0,4,6,\dots$\\
$A_2$ & 1 & $3,6,7,\dots$\\
$E$ & 2 & $2,4,5,\dots$\\
$T_1$ & 3 & $1,3,4,\dots$\\
$T_2$ & 3 & $2,3,4,\dots$\\ \hline
\end{tabular}
\caption{The table shows the single-valued irreducible representations
  $\Lambda$ of the cubic group $O$, together with their dimensions
  $d_\Lambda$ and continuum spin content $J$\protect.  Additional
  superscripts are employed to denote charge conjugation $C$ and
  parity $P$.\label{tab:irrep}}
\end{table}

Applying the projection formula \ref{eq:proj} to the operator product
of eqn.~\ref{eq:product}, we obtain
\begin{equation}
M^{\Lambda_{\alpha\beta}}_{ij} = \frac {d_\Lambda}{g_{\cal O}} \sum_{R \in O}
\Gamma_{\alpha\beta}^{\Lambda}(R) \Gamma^{\Lambda_K}(R)_{i i'} K_{i'}
\Gamma^{\Lambda_L}(R)_{j j'} L_{j'} \label{eq:proj_prod}.  
\end{equation}
Identifying a set of $d_{\Lambda}$ linearly independent operators from
the set $M^{\Lambda}_{ij}$, we obtain the Clebsch-Gordon coefficients
for obtaining the irreducible representation $\Lambda$ in the product
of operators transforming according to the irreps $\Lambda_K$ and $\lambda_L$:
\begin{equation}
M_i^{\Lambda} = Q_{ijk} K^{\Lambda_K}_j L^{\Lambda_L}_k.
\end{equation}

In Ref. \cite{Liao:2002rj}, the $\mathbb{D}$ and $\mathbb{B}$ operators are
constructed via Clebsch-Gordan coefficients of  $T_1(\nabla) \otimes
T_1(\nabla)$ to transform as $T_2$, $T_1$ irreps respectively. The
complete decomposition of $T_1 \otimes T_1$ is  $A_1 \oplus T_1 \oplus T_2 \oplus E$ with the missing $A_1$
being the laplacian and the $E$ being the $\mathbb{E}$ operator
defined in the previous section. The Clebsch-Gordan coefficients in
this case are $\mathbb{Q}_{ijk}$ where all elements are zero except
\begin{equation}
\mathbb{Q}_{111} = \tfrac{1}{\sqrt{2}};\;\; \mathbb{Q}_{122} = -
  \tfrac{1}{\sqrt{2}};\;\; \mathbb{Q}_{211} = -\tfrac{1}{\sqrt{6}};\;\; \mathbb{Q}_{222} = -\tfrac{1}{\sqrt{6}};\;\;  \mathbb{Q}_{233} = \tfrac{2}{\sqrt{3}}. 
\end{equation}

Examining the zero three-momentum lattice irrep projections of the
continuum overlaps in appendix \ref{overlaps}, we can see that certain
operators of, say, $T_1$ type can have overlap with higher spins {\em only
at finite $a$} with that overlap vanishing as $a \to 0$. The simplest
example is the local fermion bilinear $\bar{\psi} \gamma^i \psi$
which, transforming as $T_1$ can have overlap with continuum spins
$1,3,4\ldots$ at finite $a$ but can only have overlap with spin-1 in
the continuum. Several of our operators do retain overlap on to higher
spins as $a \to 0$, e.g. the $T_2$ projection of $\rho \times
\mathbb{D}$: $ (\rho \times \mathbb{D})^{T_2}_i= \epsilon_{ijk}
\bar{\psi} \gamma_j |\epsilon_{klm}| \overleftrightarrow{D}_l
\overleftrightarrow{D}_m \psi$. The non-zero overlaps in the continuum
at rest are $\langle 0 | (\rho \times \mathbb{D})^{T_2}_i |
2^{--}(\vec{0}, r) \rangle \propto |\epsilon_{ijk}|
\in^{jk}(\vec{0},r)$ and $\langle 0 | (\rho \times \mathbb{D})^{T_2}_i
| 3^{--}(\vec{0}, r) \rangle \propto \epsilon_{ijk} |\epsilon_{klm}|
\in^{jlm}(\vec{0},r)$ which are linear in the spin-2 and spin-3
Minkowski polarization vectors $\in^{\mu\nu}(\vec{p},r)$, $\in^{\mu\nu\rho}(\vec{p},r)$. None of our operators retain overlap on to spin
4 or higher in the continuum. The full set of lattice irrep
projections is displayed in appendix \ref{ops}.

\subsection{Quark and link field smearing}
\begin{figure}[h]
  \vspace{1cm}
       \psfig{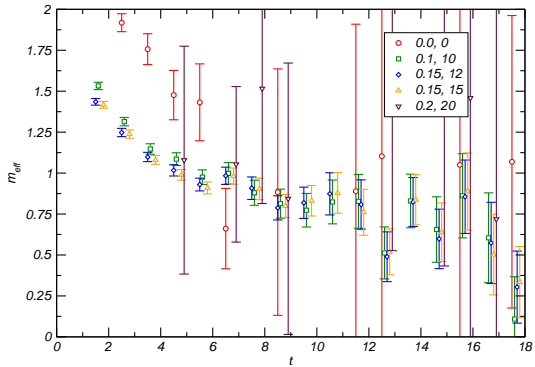}
      \caption{ Effective mass (on 300 configurations) for the
       correlator $(\rho \times
        \mathbb{B})_{T_1}-(\rho \times \mathbb{B})_{T_1}$ with link
        smearing at source and sink. Data are
        labeled by the stout-link smearing parameters, $\rho, N_\rho$.}
  \label{fig:link}
\end{figure}

Constructing covariant derivatives using finite displacements with
link fields can potentially produce rather noisy correlators. Any
individual link is subject to large UV fluctuations over the ensemble
of gauge fields. A suitable average over links neighboring a given
link can help to reduce this fluctuation and providing one does this in a
gauge-invariant and rotationally symmetric way, the quantum numbers of
the operator will not be changed. This subject is discussed in detail
in \cite{Lichtl:2006dt}. In figure \ref{fig:link} we show the reduction in
correlator noise we were able to obtain by judicious choice ($\rho =
0.15$, $N_\rho=12$) of stout-link smearing parameters.

Quark bilinear operators that are local or almost local (e.g. single site
displacements) tend to have considerable overlap with a tower of
states. While this can be helpful in a study like this one attempting
to consider excited states, it is also useful to be able to emphasize
the lower states in the spectrum. This can be achieved by smearing the
quark fields over space with a gauge-invariant cubic approximation to a rotationally symmetric gaussian with any derivatives applied subsequently,
\begin{equation}
\hspace{-5mm}  \left(1- \tfrac{3 \sigma^2}{2 N} \right)^N \left( 1 +
    \tfrac{\sigma^2/4N}{1- 3\sigma^2/2N} \sum_{i=1}^3 \left[ U_{x,i}
      \delta_{x, x+\hat{i}} + U^\dag_{x-\hat{i},i} \delta_{x,x-\hat{i}}
    \right]    \right)^N  \xrightarrow[N \to \infty]{}  e^{\sigma^2 \nabla^2/4}.\label{eq:gaussian}
\end{equation}

\begin{figure}[h]
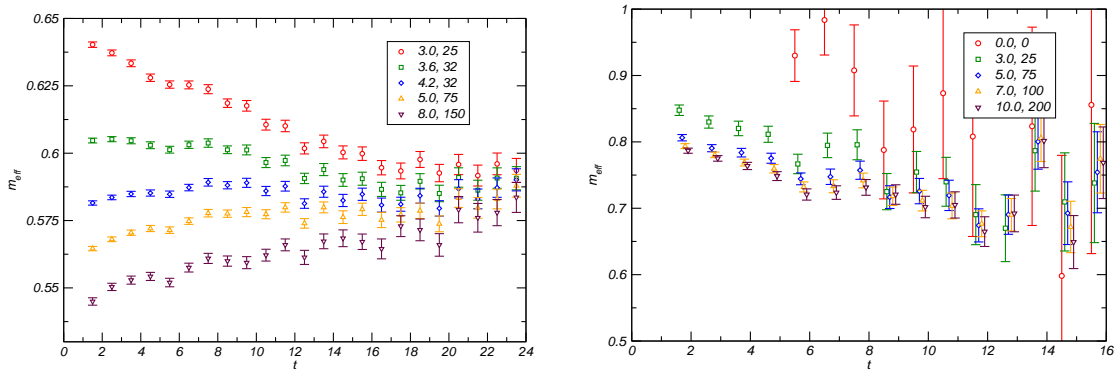

  \vspace{1cm}
       \psfig{width=7cm,file= meff_a2_rhoxNABLA_SP_quarksmear.eps}
 \hspace{.5cm}
       \psfig{width=7cm,file= meff_pi1_rhoxB_SP_quarksmear.eps}
      
      \caption{Effective masses (on 300 configurations) for the correlators $(\rho \times
        \nabla)_{T_1}-(\rho \times \nabla)_{T_1}$ and  $(\rho \times
        \mathbb{B})_{T_1}-(\rho \times \mathbb{B})_{T_1}$. Stout-link
        smearing is applied with $\rho=0.15, \,N_\rho=12$ and the data
        are labeled by the parameters of the quark smearing
        ($\sigma,\, N$) which is applied only at the source. }
  \label{fig:quark}
\end{figure}

As well as producing improved plateaus, optimizing the quark smearing
parameters does in some cases reduce noise. We show in figure
\ref{fig:quark} two representative cases.  Our final parameter selection was $\sigma =
4.2, \, N=50$ for the local, $\nabla$ and $\mathbb{D}$ operators and
$\sigma = 5.0,\, N=75$ for the $\mathbb{B}$ operators.

\section{Variational Method}\label{var}

Our spectrum results follow from application of a variational method
to matrices of correlators. This method takes advantage of the
orthogonality of state vectors on a basis of interpolating operators. The
basic numerical problem to be solved is of generalized eigenvalue type
\begin{equation}
C(t) v_\alpha = \lambda_\alpha(t) C(t_0) v_\alpha.
\end{equation}
In this expression $C(t)$ is the matrix of correlators at timeslice
$t$, i.e. $C_{ij}(t) = \langle {\cal O}_i(t) {\cal O}_j(0)
\rangle$. We transform all our operators to Minkowski space (but
retain imaginary time) and in doing so ensure that this matrix is
Hermitian (more detailed discussion of this point can be found in
Appendix \ref{ops}). The generalized eigenvectors are orthonormal on the metric
$C(t_0)$, $v_\alpha^\dag C(t_0) v_\beta = \delta_{\alpha \beta}$. The
principal correlators $\lambda_\alpha(t)$ are the generalized
eigenvalues on a given timeslice - they can be shown to behave at
large times like \cite{Variational:1985, Variational:1990}
\begin{equation}
\lambda_\alpha(t) = e^{ - m_\alpha (t - t_0)} ( 1+ {\cal O}(e^{ -
  |\delta m|
  (t - t_0)}) ),
\end{equation}  
where $m_\alpha$ is the mass of a state labeled by $\alpha$ and
$\delta m$ is the mass gap to the nearest state to $\alpha$.

Performing a Cholesky decomposition on the Hermitian matrix $C(t_0) =
LL^\dag$, where $L$ is lower diagonal, one can cast this as a conventional eigenvalue problem
\begin{equation}
L^{-1}C(t) L^{\dag -1} (L^\dag v)_\alpha = \lambda_\alpha(t) (L^\dag v)_\alpha,\label{chol}
\end{equation}
where the eigenvectors are $u_\alpha =  (L^\dag v)_\alpha$. These
eigenvectors form a $\mathrm{dim}(C) \times \mathrm{dim}(C) $ unitary
matrix that transforms our ``trial'' operators into the optimum linear
combinations to overlap with the $\mathrm{dim}(C)$ lightest states
accessible to them. The eigenvectors should be time-independent to the
extent that the correlator is saturated by  $\mathrm{dim}(C)$ states; in practice we solve the eigenvalue problem on each
timeslice and obtain eigenvectors that can vary with $t$.

Clearly from the principal correlators one can extract information
about the mass spectrum. There is also useful information in the
eigenvectors. The conventional spectral decomposition of a two-point
correlator (at zero momentum) has the form
\begin{equation}
C_{ij}(t) =  \langle {\cal O}_i(t) {\cal O}_j(0)\rangle = \sum_\alpha
\frac{Z_i^{\alpha*} Z_j^\alpha}{2 m_\alpha} e^{-m_\alpha t}, \label{spec}
\end{equation}
where the overlap factor, $Z_i^\alpha = \langle 0 | {\cal O}_i |
\alpha \rangle$. Examining the form of the generalized eigenvalue equation
with substitution of the spectral decomposition one finds that 
\begin{equation}
Z^\alpha_i = (V^{-1} )^\alpha_i  \sqrt{2 m_\alpha} e^{m_\alpha t_0 / 2}. 
\end{equation}
The inverse of the eigenvector matrix is trivial to compute owing to the
orthonormality property $V^\dag C(t_0) V = I  \implies V^{-1} = V^\dag C(t_0)$.

In practice we work with a finite space of operators and in this case
the parameter $t_0$ plays an important role. The eigenvectors are
forced by the solution procedure to be orthogonal on the metric
$C(t_0)$ - this will only be a good approximation to the true
orthogonality (which in the continuum is defined with an infinite
number of states and operators) if the
correlator at $t_0$ is dominated by the lightest $\mathrm{dim}(C)$
states. As such one should choose $t_0$ large enough that you believe
the above statement to be true. We demonstrate this using toy data in
Appendix \ref{toy}.

In choosing a $t_0$ value for a given $C(t)$ there are two factors to
take into account - the above discussion suggests we should push $t_0$
out to larger values, where the contributions of higher excited states
have decayed exponentially; however, as we do so we get into a region where
the correlator data are typically noisier. As we can see from equation
\ref{chol}, the noise on $C(t_0)$ will enter into the solution of the
eigenvalue problem at all timeslices and as such we do not want to
make $t_0$ too large. We need a criterion to decide upon an optimum
value of $t_0$ - our choice was to define a chi-squared-like quantity
gauging how well the generalized eigenvalue solution (with time-independent $Z$ values) describes the
correlators. At a given $t_0$ we solve the eigenvalue problem to
yields masses (from fits to the principal correlators - details of
the fitting follow later) and
$Z$'s. With these in hand we can reconstruct any correlator matrix
element using equation \ref{spec}. A suitable chi-squared-like
quantity can be defined as
\begin{equation}
\chi^2 = \frac{1}{\tfrac{1}{2} N (N +1 ) (t_{\mathrm{max}} - t_0) - \tfrac{1}{2}N (N +3 )
}\sum_{i, j \geq i} \sum_{t, t'=t_0+1}^{t_{\mathrm{max}}} (C_{ij}(t) -
C^{\mathrm{rec.}}_{ij}(t)) \mathbb{C}_{ij}^{-1}(t, t')  (C_{ij}(t') - C^{\mathrm{rec.}}_{ij}(t')),
\end{equation}
where $N=\mathrm{dim}(C)$ and where $\mathbb{C}$ is the data correlation matrix for the correlator
$C_{ij}$ computed with jackknife statistics. 

The optimum value of $t_0$ is chosen to be that which minimizes the
chi-squared-like quantity. In fact, since we solve the eigenvalue
problem on each timeslice we actually get $Z(t)$; we choose to take
the $Z$ values (for a given $t_0$) from a fixed timeslice $t_Z > t_0$
such that the chi-squared-like quantity is minimized at this
$t_0$. Since the $Z(t)$ are reasonably flat the chi-squared variation
with $t_Z$ is fairly mild. Insisting that the $Z$'s are time
independent for $t>t_0$ is a reflection of the fact that the only
time-dependence in the spectral representation \ref{spec} is in the exponentials.

There are some subtleties that arise because we are numerically
solving the generalized eigenvalue problem multiple times. One is that the phase of
the eigenvector for a given state is not a quantity that can be
determined with only two-point correlator information. From
equation \ref{spec}, it is clear that one could multiply
all $Z^\alpha_i$ for a fixed $\alpha$ by the same phase and equation
\ref{spec} would be unchanged. We solve the eigenvalue problem once
for each single elimination jackknife sample of our ensemble. In general the eigenproblem solver we used does
not always obtain the same phase for each jackknife sample, the
net effect of which is to make the configuration averages of the $Z$'s
appear to be much noisier than they really are. Our
fix for this is to apply a phase convention that the largest element
of the eigenvector $u = L^\dag v$ for a given state should be positive
and real.

Another subtlety can arise in the case that two (or more) states are degenerate
within the level of the configuration fluctuations. In that case the
solver may mis-assign a state label on any given jackknife sample (it orders according to the size of the
principal correlators). This is not a serious problem for the
masses since they are, in this case, equal within the error bar
anyway, but for the eigenvectors it can be
troublesome. On different jackknife samples one might see the
eigenvector flip between two orthogonal choices with the result that
the configuration average appears to be noisy. We have examined the
data in this study and found that this effect does not occur at any
meaningful level on our ensemble of 1000 configurations. We do, on occasion, see a flip of the eigenvector
between neighboring timeslices, as we show in an example below. It is for this reason that we do not choose to average over
timeslices in the determination of $Z$ from $Z(t)$.

The principal correlators are fitted using either a single exponential
or a sum of two exponentials with the constraint that $\lambda(t=t_0)
= 1$. The final timeslice used in the fit is chosen by the requirement that
the fractional error be below $10 \%$ and the first is selected by
maximizing a fit criterion. This fit criterion also decides between
the one or two exponential hypotheses.

\begin{figure}[h]
  \vspace{1cm}
       \psfig{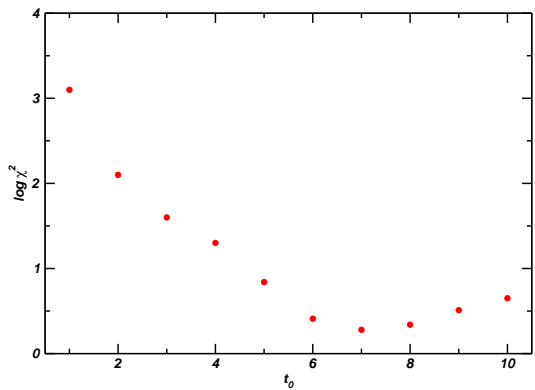}
      \caption{Chi-squared-like parameter as a function of $t_0$ for
        the $A_1^{-+}$ channel.}
  \label{fig:pion_chisq}
\end{figure}

\begin{figure}
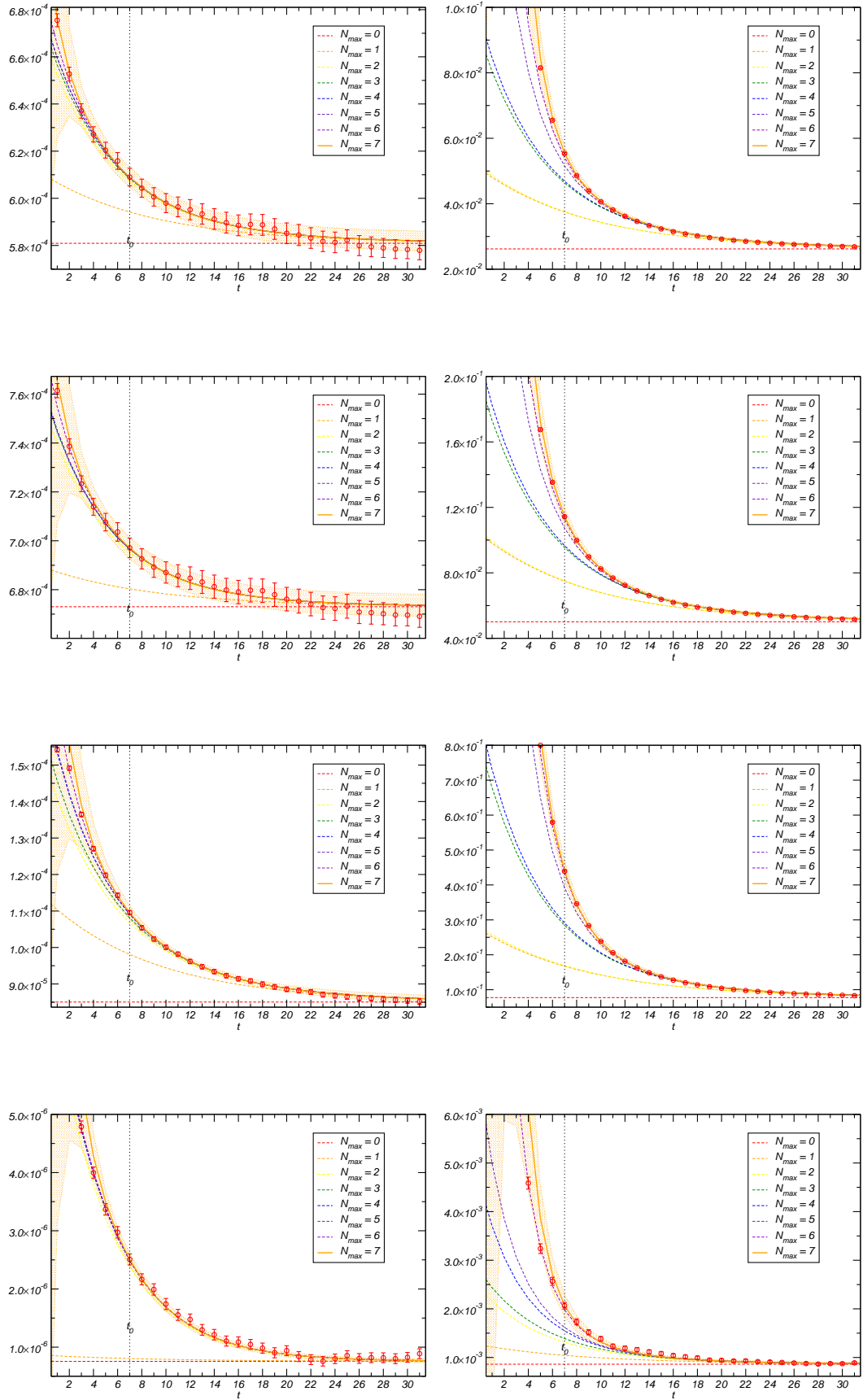

  \centering
      \psfig{width=7.0cm,file= recon_plot0_0.eps}
     \psfig{width=7.0cm,file= recon_plot1_1.eps}
      \psfig{width=7.0cm,file= recon_plot2_2.eps}
     \psfig{width=7.0cm,file= recon_plot3_3.eps}
     \psfig{width=7.0cm,file= recon_plot4_4.eps}
      \psfig{width=7.0cm,file= recon_plot5_5.eps}
      \psfig{width=7.0cm,file= recon_plot6_6.eps}
     \psfig{width=7.0cm,file= recon_plot7_7.eps}
  \caption{Diagonal correlators and variational solution
    reconstructions (ground state exponential divided out) in the case
    $t_0 =7$. Top to bottom operators are $\gamma^5,\, \gamma^0
    \gamma^5,\, b_1 \times \nabla_{A_1},\, \rho\times B_{A_1}$. Left
    column smeared, right column unsmeared. Cumulative state
    contributions are shown by the dashed lines.} 
  \label{fig:pion_recon}
\end{figure}

\begin{figure}[h]
  \vspace{1cm}
       \psfig{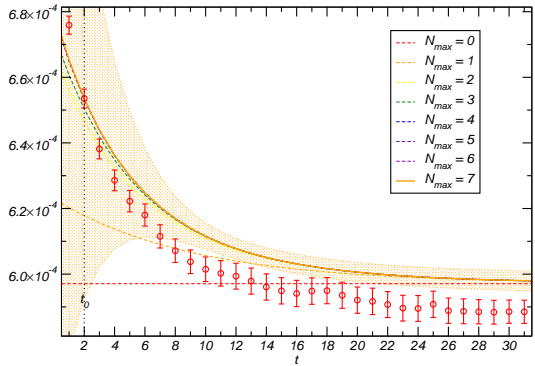}
      \caption{Example reconstructed diagonal correlator in the case $t_0 = 2$.}
  \label{fig:pion_recon_t02}
\end{figure}

As a concrete example of our solution scheme, consider the $A_1^{-+}$ channel in which we use
the following operator basis: $\big( \gamma^5|_{\mathrm{sm, us}},\, \gamma^0
\gamma^5|_{\mathrm{sm, us}},\, b_1\times\nabla_{A_1}|_{\mathrm{sm, us}},\, \rho\times
B_{A_1}|_{\mathrm{sm, us}} \big) $\footnote{Where sm and us indicate
  smeared and unsmeared, respectively}.  We solve the generalized eigenvalue problem
for all $t_0$ between $1$ and $10$ - the $\chi^2$-like parameter so
determined is shown in figure \ref{fig:pion_chisq}, with a clear minimum being observed
at $t_0=7$. We show in figure \ref{fig:pion_recon} the reconstructed diagonal
correlators obtained using the solution at $t_0 = 7$ (and $t_Z = 11$). For comparison
in figure \ref{fig:pion_recon_t02} we show a reconstructed diagonal correlator from the
solution with $t_0=2$. This variational method is only reliable if the
eigenvectors of $C(t)$ are orthogonal on the metric of $C(t_0)$, which
only occurs if $C(t_0)$ is saturated by $\rm{dim}(C)$ states. In the
$t_0 = 2$ case $C(t_0)$ is not saturated by the
eight states available\footnote{which can also be seen in the shortfall at $t < 7$
in the $t_0=7$ solution, indicating the need for additional states to
fully describe the data at all $t$ values}, and we subsequently force the eigenvectors
of $C(t)$ to be orthogonal on the ``wrong'' metric, a truncated metric belonging
to a larger Hilbert space - this shows up at
larger times as a poor description of the data. See also appendix \ref{toy}
where this effect is investigated using toy data.

Returning to the reconstructed correlators with $t_0=7$, we can see
the power of this variational method over more conventional
multi-exponential fits. Considering for example the final correlator
plotted, we see that even at timeslice 7 there are at least 6 states
contributing considerably to the correlator. It is unlikely that
a multi-exponential fitter would converge to a solution with a sum of
6 exponentials fitting over the range 7-32. There are approximate
degeneracies in the extracted spectrum and in this case the only
distinguishing feature of the states are the $Z$'s or equivalently the
eigenvectors. Without enforcing orthogonality it is hard to see how
one would extract meaningful information on these degenerate states.

\begin{figure}
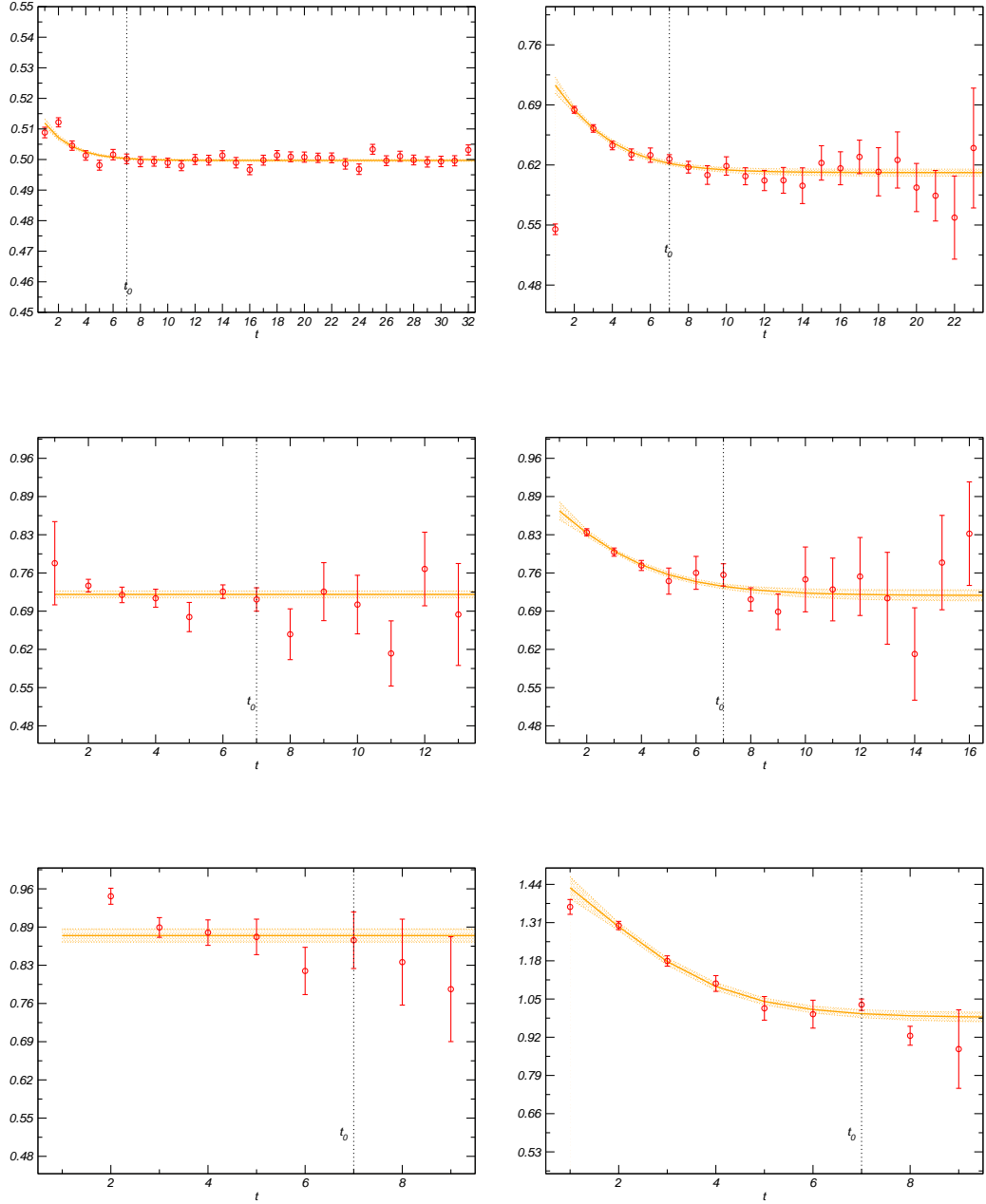

  \centering
      \psfig{width=7.0cm,file= meff_gs.eps}
     \psfig{width=7.0cm,file= meff_1st.eps}
      \psfig{width=7.0cm,file= meff_2nd.eps}
     \psfig{width=7.0cm,file= meff_3rd.eps}
      \psfig{width=7.0cm,file= meff_4th.eps}
      \psfig{width=7.0cm,file= meff_5th.eps}
  \caption{Effective masses of principal correlators and effective masses of fit solutions (one or two exponential fits)} 
  \label{fig:meff_fit}
\end{figure}

In figure \ref{fig:meff_fit} we display the fits to the principal correlators of the
lightest 6 states. Shown is the effective mass although we remind the
reader that the correlator itself is fitted.

\begin{figure}[h]
  \vspace{1cm}
       \psfig{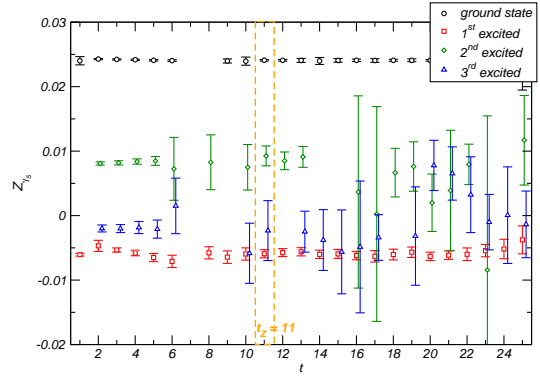}
      \caption{$Z$ values for the smeared $\gamma^5$ operator. Lowest
        four levels shown. Phases not relevant}
  \label{fig:Zpion}
\end{figure}

In figure \ref{fig:Zpion} we display the extracted $Z(t)$ values for the lightest
four states for the smeared $\bar{\psi} \gamma^5 \psi$ operator (points with very large noise have been removed). The chi-squared-like
parameter is minimized at the point $t_Z = 11$ which is seen to be
consistent with the general trend of the data. Between timeslices 19 and 20 we
see a possible example of the effect mentioned earlier that the
eigenvalues for two near-degenerate states (2nd and 3rd) can flip
assignment.

We note that there is not significant curvature for $t< t_0$ even though
in general there can be, corresponding to higher mass states having
their time-dependence placed in $Z(t)$ by the variational solver.

\section{Spectrum Results}\label{spectrum}

In this study at zero three-momentum we opt to average correlators
over diagonal directions,
\begin{eqnarray}
\overline{C}^{T_1} &\equiv& \tfrac{1}{3} \sum_{i=1}^3 C^{T_1}_{ii} \\
\overline{C}^{T_2} &\equiv& \tfrac{1}{3} \sum_{i=1}^3 C^{T_2}_{ii} \\
\overline{C}^{E} &\equiv& \tfrac{1}{2} \sum_{i=1}^2 C^{E}_{ii}.
\end{eqnarray}
It is easy to show that for the orthogonalised lattice irreps
constructed these are the only non-zero entries. In general if one
does not average in this way one will extract approximately degenerate states
corresponding to different spin-projections ($r$ in Appendix \ref{overlaps}) of a
single state. 

In the following we will discuss each $PC$ combination separately.

\subsection{$J^{++}$}
\begin{figure}[h!]
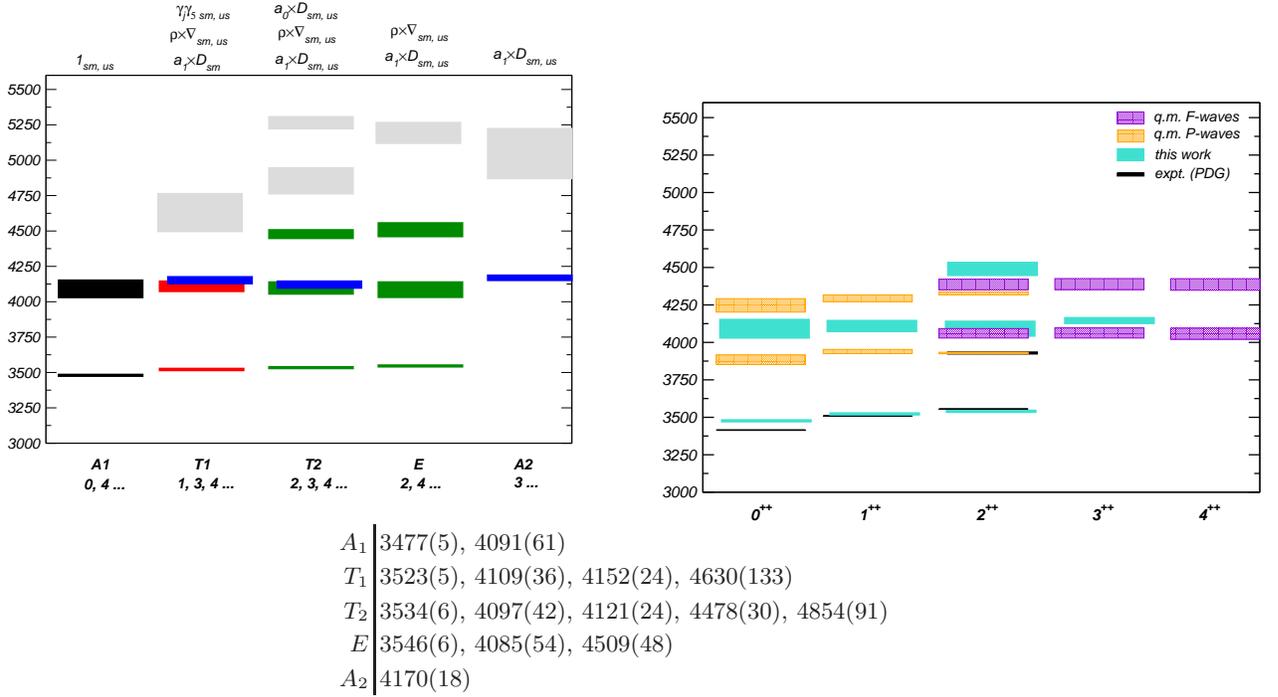

  \centering
      \psfig{width=8.0cm,file= plot++.eps}
      \hspace{1cm}
     \psfig{width=8.0cm,file= spectrum++.eps}
\begin{tabular}{r|l}
$A_1$ & 3477(5), 4091(61) \\
$T_1$ & 3523(5), 4109(36), 4152(24), 4630(133) \\
$T_2$ & 3534(6), 4097(42), 4121(24), 4478(30), 4854(91) \\
$E$ & 3546(6), 4085(54), 4509(48)\\
$A_2$ & 4170(18)
\end{tabular}
    \caption{$J^{++}$. Left pane: extracted state masses by zero-momentum
      lattice irrep. Color coding indicates continuum spin
      assignment (black=0, red=1, green=2, blue=3, grey=undetermined). Right pane:
      Comparison with experimental masses and quark potential model
      masses. Tabulated masses in MeV. } 
  \label{fig:plot++}
\end{figure}

In figure~\ref{fig:plot++} we display the states extracted. The lowest band of
states, at around 3500 MeV, can be identified with the near-degenerate
$\chi_{c0,1,2}$ states. In a potential model interpretation these
states have an internal $P$-wave and are split by a small spin-orbit interaction. In principle,
given the continuum spin content of the $A_1, T_1, T_2, E$ lattice irreps, these
states could all belong to a single $4^{++}$ state. There are a number
of reasons for disfavoring such an assignment. In the $A_1$
channel we use only the smeared and unsmeared versions of the
$\bar{\psi}\psi$ operator. In the continuum, the lowest dimension
operator that has overlap with a spin-4 meson has three covariant
derivatives - at finite lattice spacing such an operator can mix with
$\bar{\psi}\psi$, but must be suppressed by three powers of the
lattice spacing. We expect that this is sufficient suppression
relative to ${\cal O}(a^0)$ overlaps on to spin-0 states that we can
neglect it. In fact none of the operators used in the $A_1^{++}, T_1^{++}, T_2^{++},
E^{++}$ channels have any overlap with $4^{++}$ in the continuum
limit (see Appendix \ref{overlaps}), but the finite $a$ overlap is not always as suppressed as in
the point-like case.

Support beyond simply mass degeneracy for the common $2^{++}$ assignment of the lightest state in
$T_2$ and $E$ comes from the extracted $Z$-values. Consider for
example the operator $\rho \times \nabla$ - this has overlap in the continuum
limit, at zero three-momentum, on to a $2^{++}$ state as follows:
\begin{equation}
  \langle 0 | \psi \gamma^i D^j \psi | 2^{++}(\vec{0}, r)\rangle = Z \in^{ij}(\vec{0},r).
\end{equation}
Now, while we might think of the $T_2$ and $E$ irreps as being
independent on a discrete lattice, their particle content is clearly related in the continuum limit,
e.g. they share a common $Z$-value:
\begin{eqnarray}
 \langle 0 | (\rho \times \nabla)_{T_2}^i  |2^{++}(\vec{0},r)\rangle &=
 |\epsilon_{ijk}| \langle 0 | \psi \gamma^j D^k \psi | 2^{++}(\vec{0},
 r)\rangle &= Z |\epsilon_{ijk}|  \in^{jk}(\vec{0},r)\\
 \langle 0 | (\rho \times \nabla)_{E}^i  |2^{++}(\vec{0},r)\rangle &=
 \mathbb{Q}_{ijk} \langle 0 | \psi \gamma^j D^k \psi | 2^{++}(\vec{0},
 r)\rangle &= Z \mathbb{Q}_{ijk}  \in^{jk}(\vec{0},r).
\end{eqnarray}
We might reasonably expect that if our simulation can be considered to
be ``close'' to the continuum limit, the $Z$-values extracted from
the $T_2$ and $E$ channels would be related as above, up to hopefully
small corrections in powers of $a$. This is in fact
what we find to a high accuracy. As a result of the correlator
direction averaging described above, the relation of the extracted $Z$'s
to the $Z$ defined above is $Z_{T_2} = \sqrt{2} Z;\; Z_E = Z$. For the
lightest state in $T_2,E$ we find for the smeared operator that $\tfrac{Z_{T_2}}{\sqrt{2} Z_E}
= 1.00(1)$. We take this as evidence that these states are rather
close to being components of the same $2^{++}$ state.

In addition, for the unsmeared $\rho \times \nabla$ operators we find $\tfrac{Z_{T_2}}{\sqrt{2} Z_E}
= 1.00(3)$. An equivalent analysis can be applied to the $a_1 \times
\mathbb{D}$ operator yielding $\tfrac{Z_{T_2}}{\sqrt{2} Z_E}
= 0.99(3)$ for smeared and $\tfrac{Z_{T_2}}{\sqrt{2} Z_E}
= 1.12(5)$ for unsmeared, all of which appear to be compatible with a
common state assignment.

The second band of states (around 4100 MeV) features a state in the
$A_2$ irrep, which we associate with a spin-3 state in the continuum
(neglecting spin-6 or higher possibilities). This state should be
partnered by nearby states in the $T_1$ and $T_2$ irreps for which
there are candidates being either the first or second excited states
in each channel. We can use the fact that in the continuum, the $T_2$
and $A_2$ projections of the operator $a_1 \times \mathbb{D}$ share a
common $Z$-overlap onto the spin-3 state:
\begin{eqnarray}
 \langle 0 | (a_1 \times \mathbb{D})_{A_2}  |3^{++}(\vec{0},r)\rangle &=
 |\epsilon_{ijk}| \langle 0 | \psi \gamma^j \gamma^5 \mathbb{D}^k \psi | 3^{++}(\vec{0},
 r)\rangle &= Z |\epsilon_{ijk}|  \in^{ijk}(\vec{0},r)\\
 \langle 0 | (a_1 \times \mathbb{D})_{T_2}^i  |3^{++}(\vec{0},r)\rangle &=
 \epsilon_{ijk} \langle 0 | \psi \gamma^j \gamma^5 \mathbb{D}^k \psi | 3^{++}(\vec{0},
 r)\rangle &= Z \epsilon_{ijk} |\epsilon^{klm}| \in^{jlm}(\vec{0},r).
\end{eqnarray}
With the direction averaging used we find that $Z_{A_2} = \sqrt{6} Z$
and $Z_{T_2} = \sqrt{\tfrac{8}{3}} Z$. The extracted $Z$-values for
the smeared operator give
\begin{equation}
  \frac{2 Z_{A_2}(\mathrm{gnd. st.})}{3 Z_{T_2}(\mathrm{1^{st} ex.})}
  = 0.15(15);\;\;
\frac{2 Z_{A_2}(\mathrm{gnd. st.})}{3 Z_{T_2}(\mathrm{2^{nd} ex.})}  = 0.97(9).
\end{equation}
Hence it would appear that the $\mathrm{2^{nd}}$ excited state in the
$T_2$ channel should be partnered with the ground state in $A_2$ as
components of a spin-3 meson. The remaining components should reside
in the $T_1$ channel. In fact none of the operators used in that
channel have overlap with $3^{++}$ in the continuum, but overlap can
occur at finite $a$ through mixing with continuum operators of equal
or higher dimension. For example, we have included an operator $(a_1
\times \mathbb{D})_{T_1}$ which can mix with the same-dimension
operator $(a_1 \times \mathbb{E})_{T_1}$ which does have overlap with
$3^{++}$ in the continuum. The degree of mixing is presumably
related to a power of $g(a)$ with at most logarithmic divergence with
$a$. At this stage we will assume that
one of the states around 4100 in $T_1$ is a component of this spin-3
state.

The remaining states in this band then can either be components of a
spin-4 meson or a nearly degenerate $0^{++}, 1^{++}, 2^{++}$ set. As
before an argument against the $4^{++}$ option is that none of the
operators used has overlap with spin-4 in the continuum - while this
is likely to be a strong constraint in the $A_1$, $\bar{\psi}\psi$
case, it is less convincing in the other channels where we include
operators featuring two derivatives. Here the spin-4 overlap may only
be suppressed by one power of $a$. Explicitly we find that the overlap
of the unsmeared $\bar{\psi}\psi$ operator onto the first excited
state in $A_1$ is of the same magnitude as the overlap onto the ground
state.
 
We can take the alternate route of looking for support for the
$0^{++}, 1^{++}, 2^{++}$ hypothesis. We have already seen that there
should be a definite relation between the $Z$'s in $E$ and $T_2$ if
the state corresponds to components of a spin-2 meson. From the fact that we have
already assigned the $\mathrm{2^{nd}}$ excited state in the
$T_2$ channel to a spin-3 meson, we'd expect that the
$\mathrm{1^{st}}$ excited states in $T_2$ and $E$ would have $Z$
matching. In fact we find
\begin{equation}
\frac{Z_{T_2}}{\sqrt{2} Z_E}(\rho \times \nabla; \mathrm{sm}) =
0.69(28);\;\;\frac{Z_{T_2}}{\sqrt{2} Z_E}(\rho \times \nabla;
\mathrm{us}) = 0.95(27);\;\; \frac{Z_{T_2}}{\sqrt{2} Z_E}(a_1 \times
\mathbb{D}; \mathrm{sm}) = 0.58(57);\;\; \frac{Z_{T_2}}{\sqrt{2} Z_E}(a_1 \times \mathbb{D}; \mathrm{us}) = 0.60(24).
\end{equation} 
While not as convincing as the ground state, we do at least see that these numbers
are not inconsistent with the spin-2 hypothesis.

Giving spin assignments to higher states becomes increasingly difficult. It
is no longer possible to find complete sets of components of a given
spin across lattice irreps, since in certain channels we have insufficient
operators to extract further states (e.g. only two operators in $A_1$
and hence only two states). Additionally both masses and $Z$'s get
noisier as we go higher in the spectrum. We can {\em tentatively} assign the
states at around 4300 MeV in $T_2$ and $E$ to a $2^{++}$ on the basis
of their $Z(\rho \times \nabla; \mathrm{us})$ where $\tfrac{Z_{T_2}}{\sqrt{2} Z_E}
= 0.98(17)$.

We summarize our assignments in figure~\ref{fig:plot++} where they are compared to
experimental charmonium states and the potential model states of
\cite{Barnes:2005pb}. Discussion will follow in the next section.

\subsection{$J^{--}$}

\begin{figure}[h!]
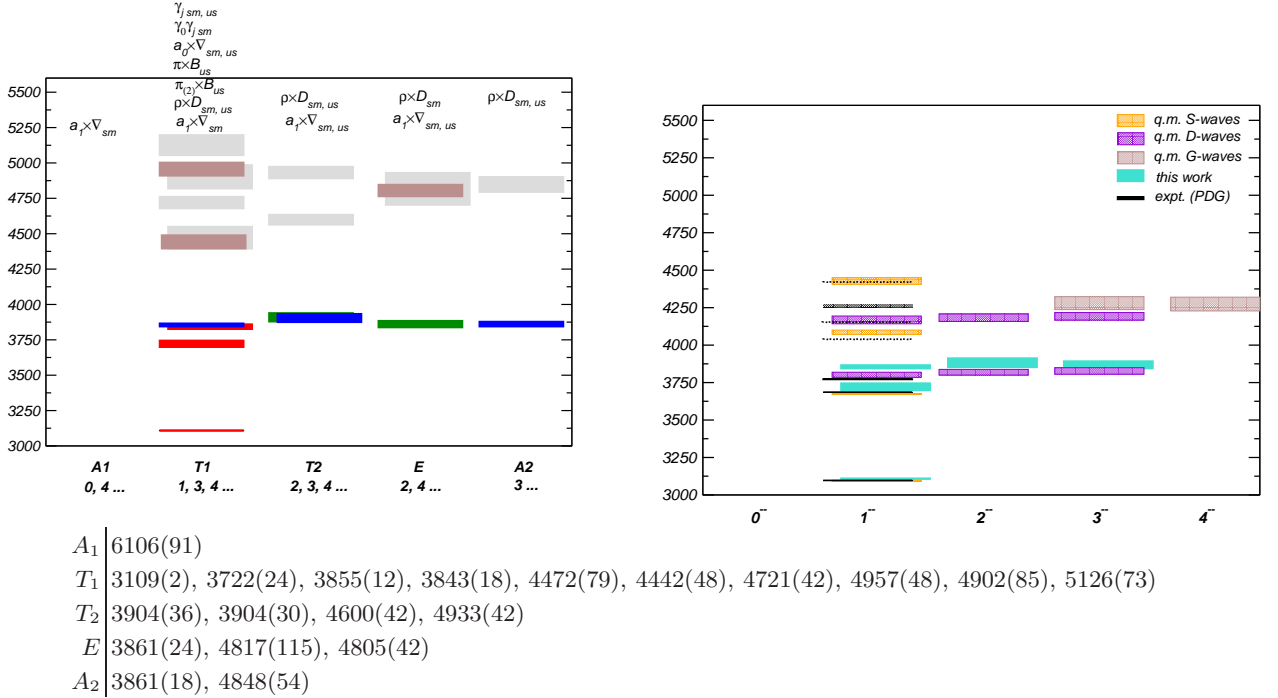

  \centering
      \psfig{width=8.0cm,file= plot--.eps}
      \hspace{1cm}
     \psfig{width=8.0cm, file= spectrum--.eps}
\begin{tabular}{r|l}
$A_1$ & 6106(91)\\
$T_1$ & 3109(2), 3722(24), 3855(12), 3843(18), 4472(79), 4442(48), 4721(42), 4957(48), 4902(85), 5126(73)\\
$T_2$ & 3904(36), 3904(30), 4600(42), 4933(42)\\
$E$ & 3861(24), 4817(115), 4805(42)\\
$A_2$ & 3861(18), 4848(54)
\end{tabular}
    \caption{$J^{--}$. Left pane: extracted state masses by zero-momentum
      lattice irrep. Color coding indicates continuum spin
      assignment (black=0, red=1, green=2, blue=3, grey,brown=undetermined). Right pane:
      Comparison with experimental masses and quark potential model
      masses. Tabulated masses in MeV.} 
  \label{fig:plot--}
\end{figure}

Results are displayed in figure~\ref{fig:plot--}. The isolated lightest state in
the $T_1$ channel can be identified as the $J/\psi$. In this channel, there is then a
large gap to the first excited state which is close to two nearly
degenerate states. We anticipate the explanation of these levels as
being two $1^{--}$ states and a $3^{--}$ state. In this case we
believe that this is the first time that a lattice QCD calculation has
observed something like the experimental $\psi(3686)/\psi(3770)$
pair. The extraction of such nearly degenerate states is made possible
by the orthogonality properties of the variational method.

The band of states around 3800 MeV has a natural explanation as a
nearly degenerate set containing a $3^{--}$ state, a $2^{--}$ state
and two $1^{--}$ states. The first excited state in the $T_2$ channel
is associated with the ground state in $E$ as a $2^{--}$ meson using
the $Z$-relation since $\tfrac{Z_{T_2}}{\sqrt{2} Z_E}(a_1 \times
\nabla); \mathrm{sm}) = 1.03(13)$.

The ground state in $A_2$ can be associated with the ground state in
$T_2$ on the basis of mass degeneracy. The $Z$-relation is at the
three-sigma edge of validity,  $\frac{2 Z_{A_2}}{3 Z_{T_2}}
  = 0.76(8)$. It is quite possible that this discrepancy is a
  discretisation effect, after all the $Z$-relations are only supposed
  to hold precisely in the continuum limit. To complete the $3^{--}$
  we associate one of the levels in $T_1$ - since none of the $T_1$
  operators we used have overlap with $3^{--}$ in the continuum we cannot
  apply any $Z$ analysis here, although we do expect to get overlap
  through mixing with $(\rho \times
  \mathbb{E})_{T_1}$ which has overlap with $3^{--}$ in the continuum.

A small overlap onto the unsmeared $\bar{\psi}\gamma_j\psi$ operator
might be taken as signal for spin-3 nature owing to the suppression of
such an overlap at ${\cal O}(a^3)$, however there are also good
continuum physics reasons why a spin-1 meson might have a small value
of this overlap. In the non-relativistic limit this overlap measures
the wavefunction at the origin of a quark-antiquark pair - while this
can be considerable for an $S$-wave state, it is zero for a $^3D_1$
state. Relativistic corrections convert this zero to a suppression
relative to $S$-wave states, see for example the experimental
$f_{\psi(3686)}(2^3S_1) = 279(8) \mathrm{MeV}$ and
$f_{\psi(3770)}(^3D_1?) = 99(20) \mathrm{MeV}$.

We find that the ground state and first excited state in $T_1$ have
comparable decay constants ($463(8)$ MeV and $416(73)$ MeV) while the second and third excited states
have decay constants consistent with zero ($182(211)$ MeV and $40(153)$
MeV). This could be explained if
they are dominantly $^3D_1$ and $^3D_3$ states. This could be
investigated with simulations at smaller $a$ where we would expect one
$f$ to remain finite (but small) while the other went to zero\footnote{provided
  we included an operator with continuum overlap on to $3^{--}$,
  otherwise the state would decouple altogether}.

It is worth noting that these decay constants\footnote{See
  \cite{Dudek:2006ej} for details of scale setting of such quantities on
  an anisotropic lattice} are larger than the experimental values. A
similar result was seen for the ground state in the preliminary $N_F =
2+1$ results of \cite{diPierro:2003bu}. We are somewhat surprised
that this value is so high considering the relative experimental agreement using
domain wall fermions on the same lattices in \cite{Dudek:2006ej}. We
might propose that since we are not improving our vector current
operator we cannot expect the full ${\cal O}(a)$ Clover improvement, however in
the $m a_s$ improvement scheme proposed in \cite{Harada:2001ei} there is no
improvement at $\vec{p}=(000)$.

We have not been able to make spin assignments for the higher states
in the spectrum for the reasons outlined in the $J^{++}$ section
above. The only $A_1^{--}$ state we were able to extract is rather
close to our temporal cutoff scale (and is compatible with the mass
reported in \cite{Liu:2005rc}) - such a state could be an exotic
$0^{--}$, but could equally well be a non-exotic $4^{--}$. We will not
comment upon it further. 

We summarize our assignments in figure~\ref{fig:plot--} where they are compared to
experimental charmonium states and the potential model states of
\cite{Barnes:2005pb}. Note that we only show those states for which
we are confident of the spin assignment, the reader may assign the
higher states are their own risk.

\subsection{$J^{+-}$}

The $J^{+-}$ sector has the interesting property that all states in
with $J$-even are exotic in the sense of being inaccessible to a
fermion-antifermion bound state. Such states can be constructed from
higher Fock states and as such are often described as being
``multiquarks'' (extra quark degrees of freedom)  or ``hybrids'' (extra
gluonic degrees of freedom). Adding an extra pair of charm quarks
would take the state mass up to around 6 GeV, which is at the scale of
our cutoff and where our quenched non-unitarity might be felt. In the physical spectrum it may be possible for
light-quarks to play a non-trivial role, in this quenched study we can
say nothing about this possibility. If a non-trivial gluonic field
produces exotic quantum numbered states we have hope of seeing it
here.
\begin{figure}[h!]
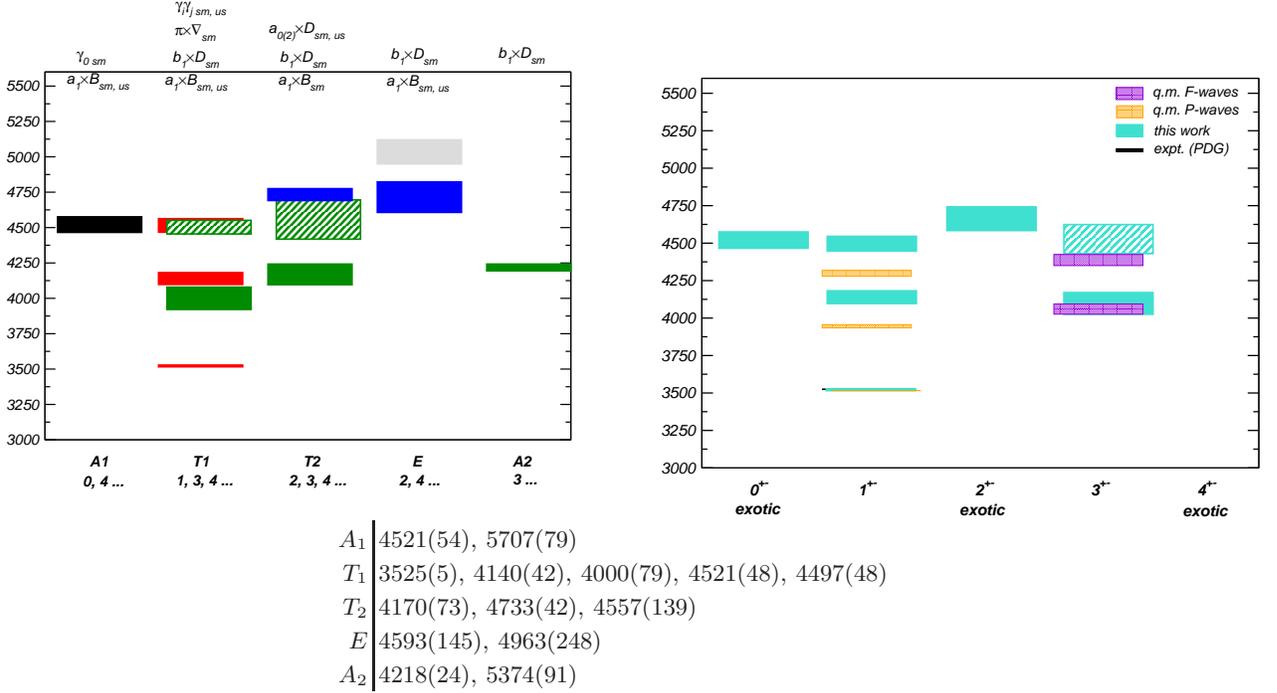

  \centering
      \psfig{width=8.0cm,file= plot+-.eps}
      \hspace{1cm}
     \psfig{width=8.0cm, file= spectrum+-.eps}
\begin{tabular}{r|l}
$A_1$ & 4521(54), 5707(79)\\
$T_1$ & 3525(5), 4140(42), 4000(79), 4521(48), 4497(48)\\
$T_2$ & 4170(73), 4733(42), 4557(139)\\
$E$ & 4593(145), 4963(248)\\
$A_2$ & 4218(24), 5374(91)
\end{tabular}
    \caption{$J^{+-}$. Left pane: extracted state masses by zero-momentum
      lattice irrep. Color coding indicates continuum spin
      assignment (black=0, red=1, green=2, blue=3, grey=undetermined). Right pane:
      Comparison with experimental masses and quark potential model
      masses. Hatching indicates that the spin assignment is not based
      upon a $Z$-analysis. Tabulated masses in MeV.} 
  \label{fig:plot+-}
\end{figure}

Our results are displayed in figure~\ref{fig:plot+-}. The lightest state in $T_1$
can be taken to be the $1^{+-}$ $h_c$ since it has no obvious partners
in other lattice irreps. We assign the set of states around 4000 MeV
to a $1^{+-}, 3^{+-}$ pair, support for the $T_2, A_2$ pairing comes
from analysis of the overlaps where $\tfrac{2Z_{A_2}}{3Z_{T_2}}(b_1
\times \mathbb{D}; \mathrm{sm}) = 1.07(7)$. Possible support for the
lighter of the $T_1$ states being a $3^{+-}$ component comes from the
overlap on the $\bar{\psi} \gamma_i \gamma_j\psi$ operator which is
consistent with zero as might be expected for an ${\cal O}(a^2)$
suppression. The other state has an overlap which is rather similar to
the ground state and we propose that it is a $1^{+-}$.

The states around 4500 MeV are rendered somewhat non-trivial by our
operator-limited knowledge of the $A_2$ channel. There might be other
spin-3 states in this channel, but without more operators we cannot
tell. If there where to be another state near 4500 MeV in $A_2$ we
would have two plausible explanations of the states, either $3^{+-},
4^{+-}$ or $3^{+-}, 2^{+-}, 1^{+-}, 0^{+-}$. We prefer the second
choice and have some support for this from the overlaps. The pairing
shown in blue satisfies $\tfrac{Z_{T_2}}{\sqrt{2}Z_{E}}(a_1
\times \mathbb{B}; \mathrm{sm}) = 1.03(10)$ and we identify it with a
$2^{+-}$. The higher state in $T_1$ have a large overlap onto
$\bar{\psi} \gamma_i \gamma_j\psi$ and we propose it is $1^{+-}$,
while the lower is consistent with zero so we tentatively assign it to
$3^{+-}$, partnered by the lower state in $T_2$ and a missing state in
$A_2$. The remaining state in $A_1$ is then proposed to be a $0^{+-}$.

Our spin-assigned spectrum is compared with experiment and quark
models in figure~\ref{fig:plot+-}. The exotic $0^{+-}, 2^{+-}$ states are
somewhat lighter than the signals reported in \cite{Liao:2002rj, Liu:2005rc}.

\subsection{$J^{-+}$}

The $J^{-+}$ sector again houses exotics, this time in $J$-odd
channels. We can safely assign the lowest two states in $A_1$ to
$0^{-+}$. The lightest states in $T_2, E$ appear to make up a
$2^{-+}$, and $Z$ comparison appears to confirm this: $\tfrac{Z_{T_2}}{\sqrt{2}Z_{E}}(b_1
\times \nabla; \mathrm{sm}) = 1.07(8)$. Above this spin assignment
becomes treacherous. We will work under the assumption that the
unexplored $A_2$ channel is in fact empty in our mass range -
$3^{-+}$ states are exotic and of high spin and might reasonably be
expected to be heavy. With this we are left with two possible
assignments of the states near 4300 MeV. The first would be to have
two nearly degenerate $0^{-+}$ states, an exotic $1^{-+}$ state and a
$2^{-+}$ state. The second possibility is to have a single $0^{-+}$
and a $4^{-+}$. Analysis of the $Z$'s in $T_2, E$ indicates the spin-2
interpretation is at the borderline three sigma level: $\tfrac{Z_{T_2}}{\sqrt{2}Z_{E}}(\rho
\times \mathbb{B}; \mathrm{sm}) = 0.78(7)$. The overlap of $A_1$
states onto $\bar{\psi} \gamma_5 \gamma_0 \psi$ is large for the
ground state, the first excited state and the third excited state, but
is consistent with zero for the second excited state. This might be
evidence for a spin-4 nature for this state, with the overlap
suppressed at ${\cal O}(a^3)$.
\begin{figure}[h!]
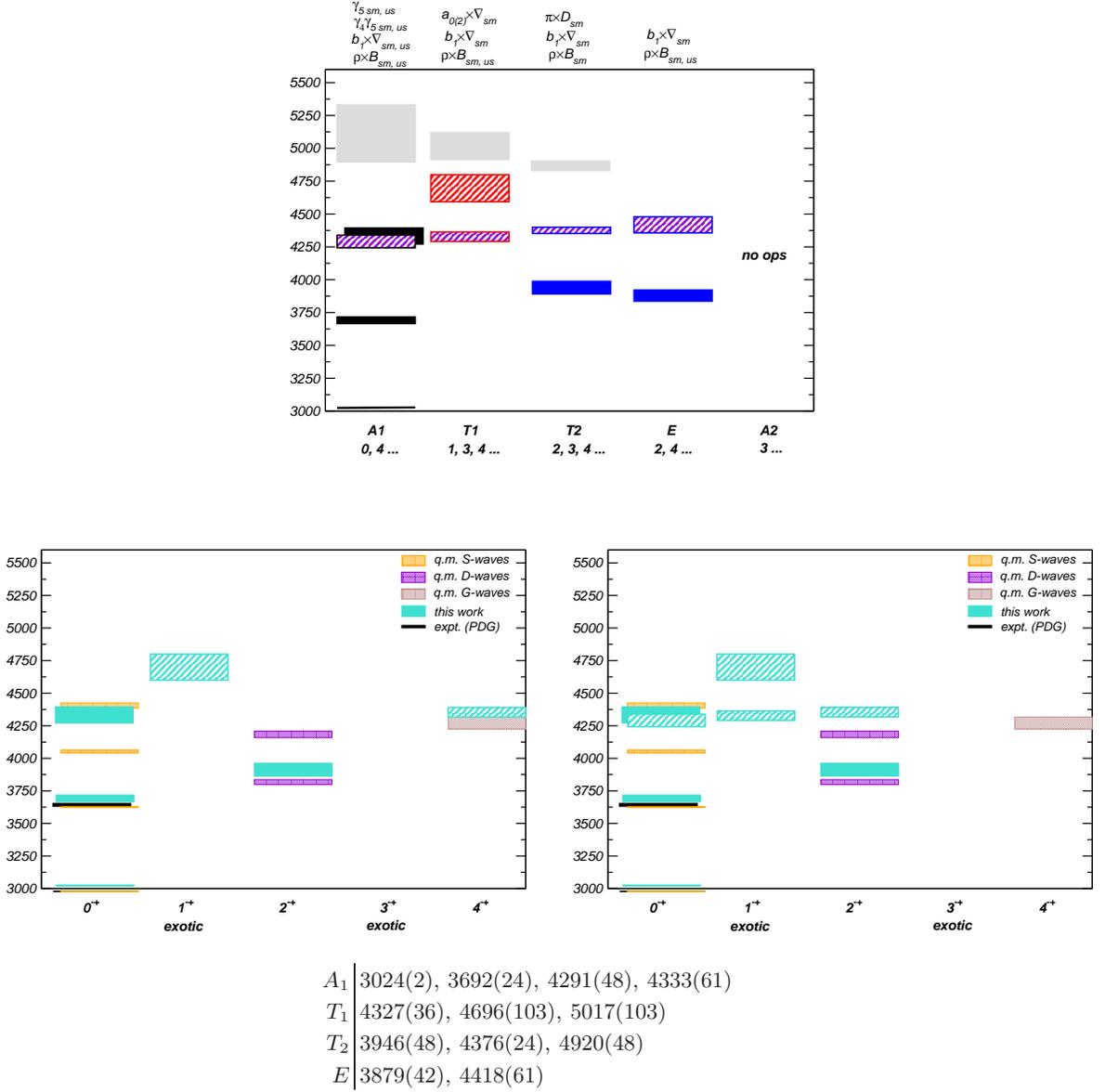

  \centering
      \psfig{width=8.0cm,file= plot-+.eps}
     \hspace{10cm}
     \psfig{width=8.0cm, file= spectrum-+.eps}
\psfig{width=8.0cm, file= spectrum-+_2.eps}
\vspace{-1cm}
\begin{tabular}{r|l}
$A_1$ & 3024(2), 3692(24), 4291(48), 4333(61)\\
$T_1$ & 4327(36), 4696(103), 5017(103)\\
$T_2$ & 3946(48), 4376(24), 4920(48)\\
$E$ & 3879(42), 4418(61)
\end{tabular}
\vspace{1cm}
    \caption{$J^{-+}$. Top pane: extracted state masses by zero-momentum
      lattice irrep. Color coding indicates continuum spin
      assignment (black=0, red=1, green=2, blue=3, grey=undetermined). Lower panes:
      Comparison with experimental masses and quark potential model
      masses. Hatching indicates that the spin assignment is not based
      upon a $Z$-analysis. Lower left pane shows the $4^{-+}$ lighter
      hypothesis. Lower right pane shows the $1^{-+}$ lighter
      hypothesis. Tabulated masses in MeV} 
  \label{fig:plot-+}
\end{figure}

We are not able with the information we have to decisively state which
of the two hypothesis above is correct, and as such in figure~\ref{fig:plot-+} we show
two possible spectrum interpretations of our data.

\begin{figure}[h]
  \centering
      \psfig{width=8.0cm,file= 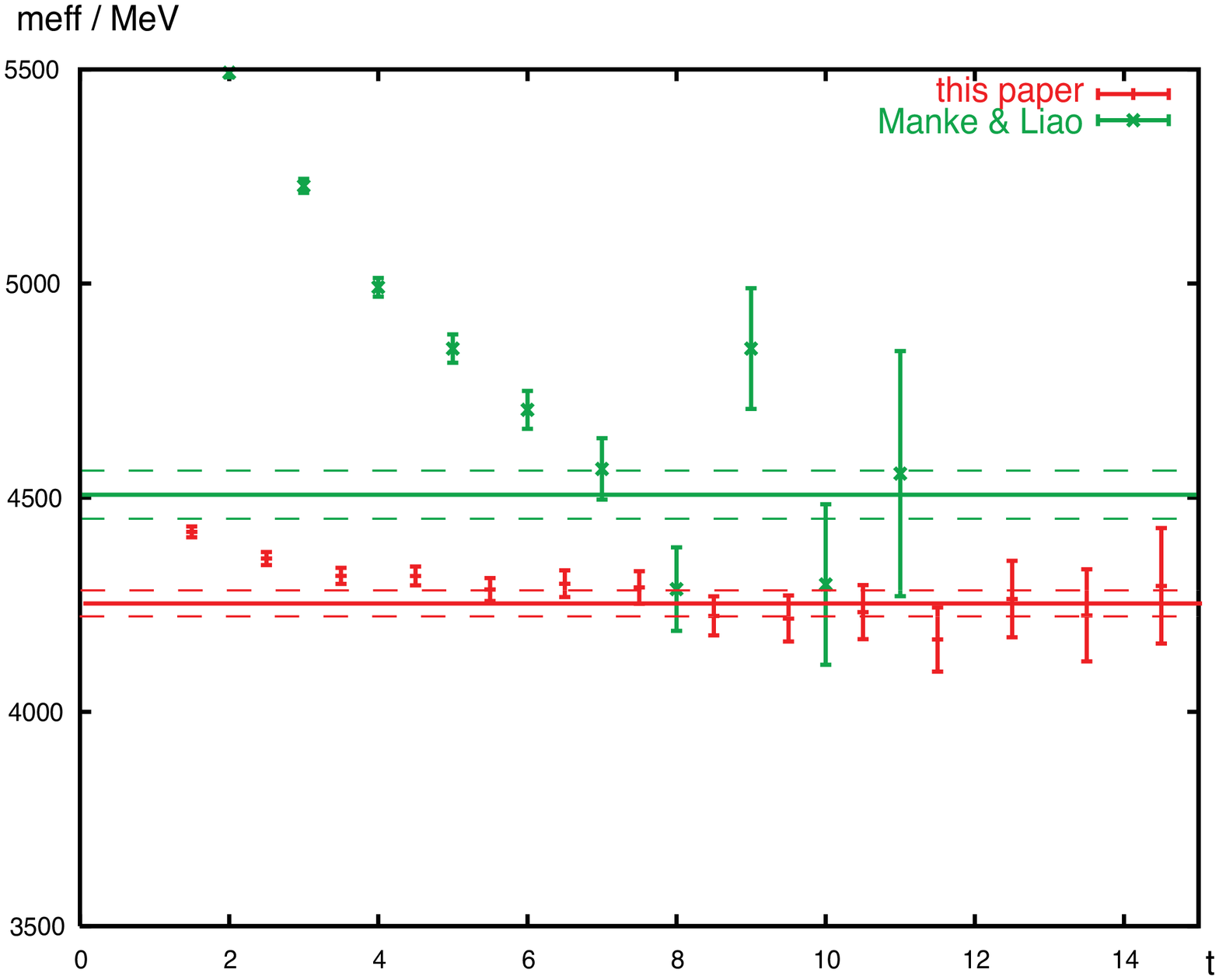}
    \caption{Effective mass of $(\rho\times B)_{T_1}$ correlator from
      this analysis and from \cite{Liao:2002rj}. Upper horizontal line is
      result of fit performed in \cite{Liao:2002rj}.} 
  \label{fig:manke}
\end{figure}

In figure \ref{fig:manke} we show the effective mass of our smeared $(\rho\times
B)_{T_1}$ correlator along with the correlator taken from \cite{Liao:2002rj}
computed on lattices with the same $a_s$ but with different anisotropy
($\xi = 2$). It is notable that the data are consistent although the
fits are not.

\subsection{Discussion}
In our results we see thatthe masses of many excited states and states of
higher spin are high with respect to the quark
potential model and, where available, experimental states. We can
identify three possible reasons for the systematic difference which
are the three approximations in our lattice study: finite volume,
non-zero $a$ and the quenched approximation\footnote{and, in
  principle, the effect of disconnected contributions to the correlators}.. 

\begin{figure}[h]
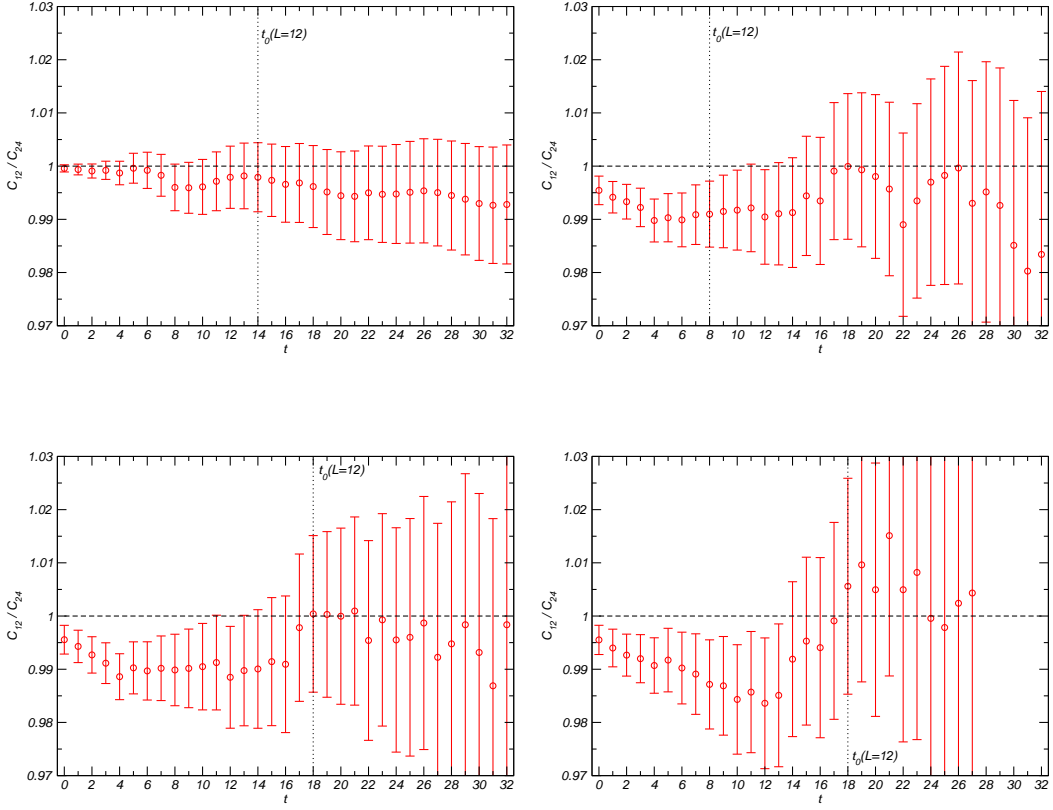

  \centering
      \psfig{width=7.0cm,file= rho-gamma_PP.eps}
   \psfig{width=7.0cm,file= a1-rhoxNABLA_SP.eps}
   \psfig{width=7.0cm,file= a2-rhoxNABLA_SP.eps}
   \psfig{width=7.0cm,file= rho2-a1xNABLA_SP.eps}
    \caption{Ratio of correlators for volumes $12^3, 24^3$. Unsmeared
      $\gamma_i-\gamma_i$, smeared-unsmeared $(\rho \times \nabla)_{T_1}$,
      smeared-unsmeared $(\rho \times \nabla)_{T_2}$, smeared-unsmeared $(a_1
      \times \nabla)_{T_2}$. Also shown are the optimum $t_0$ found in the
      $12^3$ analysis.} 
  \label{fig:12vs24}
\end{figure}

The effects of a finite volume of $(1.2\,\rm{fm})^3$ would seem to be
a likely culprit. Within potential models the wavefunction of
increasingly excited and higher spin states gets support at larger and
larger distances. Estimates from the potential model of Barnes,
Godfrey \& Swanson~\cite{Eric-private} indicate that for example the rms
radius of the first excited $\chi_{c0}$ is already over $1\,\rm{fm}$. In order to consider this possibility we computed a
limited set of correlators on a quenched $24^3 \times 48$ lattice with
all parameters identical to the simulation on $12^3 \times 48$. In
figure \ref{fig:12vs24} we show the ratio of these correlators in several typical
cases where it is clear that there is no statistically significant
difference, particularly in the $t > t_0$ region. To be clear we are
unable to see any relevant finite-volume differences
between the $(1.2\,\rm{fm})^3$ box and the $(2.4\,\rm{fm})^3$ box. 

We did not study multiple lattice spacings but we can consider results
in the literature for the $a$-dependence of state masses. The studies in
\cite{Okamoto:2001jb} and \cite{Liao:2002rj} use rather similar actions to ours on
quenched lattices and see $a$-dependences that are relatively mild if
the action is truly improved so that an extrapolation in $a_s^2$ can
be performed. If a non-negligible ${\cal O}(a_s)$ term remains the
dependence may be more significant. There typically appears to be a
decrease in state masses as the continuum is approached which may
explain part of our high masses.

In comparing to the quark-potential model the effect of thresholds can
be neglected since their effect does not appear in this model. Hence
we might expect that the principal effect of the quenched
approximation would be in the incorrect running of the coupling and
the associated scale-setting ambiguity. The ambiguity in scale-setting
can be clearly seen in the three graphs of figure 13 in
\cite{Okamoto:2001jb}. Since the wavefunctions of higher and higher excited
states sample larger distances we might expect to feel even larger
effects in our study. We suspect that quenching may be a major
contributor to our systematically high masses.

\section{Conclusions}

We have investigated the use of a large basis of interpolating fields
on the extraction of excited charmonium meson states in many $J^{PC}$
channels. We propose an application of a variational method that
systematically selects the parameter $t_0$ to ensure the best possible
description of the data at all times greater than $t_0$. We use the
eigenvalues (principal correlators) to determine the mass spectrum and
the eigenvectors to determine the overlaps of our interpolating fields
on to the states extracted. These numbers are used herein to aid
continuum spin assignment and are required quantities for study of
three-point functions with the aim of extracting transition matrix elements.

The variational method has the important feature of using the
orthogonality in a space of interpolating fields of different states -
this is a powerful lever-arm in the extraction of near-degenerate
states. In principle the method can miss states that appear in the
spectrum if no linear combination of the interpolating fields used has
a sufficient overlap on to the state in question; however, this is
equally true of any ``fitting'' method and simply suggests the use of
the broadest set of interpolating fields possible. With perfect
statistics this analysis method should not produce any ``false'', additional states
in the spectrum as can happen in simple non-linear exponential fitting
schemes where one adds states to minimize a chi-squared. With finite
statistics and a simple ordering of eigenvalues there is the
possibility that configuration-by-configuration there is state
mis-assignment which can appear as several levels appearing to be
degenerate within large errors. We did not observe effects like this on
our sample of 1000 configurations, but with a smaller ensemble and the
consequent larger fluctuations of the jackknife bins it might be
expected to occur. The effect would be partnered by large
fluctuations, configuration-by-configuration, of the eigenvectors, and
a possible mechanism to control it would be to order the levels not
simply by the size of the eigenvalue but by the similarity (via
a dot-product, say) to some
established basis of eigenvectors (for example comparing each
single-elimination bin to the eigenvectors on the
entire ensemble average).

Our interpolating fields are designed so that they have relatively
simple and known overlap structures in the continuum limit, this was
used to aid in the continuum spin identification of
states. Supplemented with analysis at more than one lattice spacing we
believe this will remove much of the possible spin ambiguity of
working on a cubic lattice. 

We observe in our results that our excited state masses are systematically
high with respect to experiment (where it is measured) and to the
potential models of \cite{Barnes:2005pb}. Our best guess is that this is a
combination of not having extrapolated to the continuum limit and the
effect of the quenched approximation. Both these issues can be
remedied with further computation now that the efficacy of the fitting
model has been demonstrated.

Arguably the most intriguing channel considered in this study is
$T_1^{-+}$. The lowest spin contributing to this is the exotic
$1^{-+}$ which has been the subject of several analyses
\cite{Juge:2006fm, Mei:2002ip, McNeile:1998cp, Manke:1998qc, Liu:2005rc}. In this study we found that we could not determine
whether the lightest state in this channel (at $\sim 4300$ MeV) is
exotic spin-1 or non-exotic spin-4. In the potential models of
\cite{Barnes:2005pb}, a non-exotic $4^{-+}$ state is expected in this mass region.  With the uncontrolled
systematics related to quenching in this work we can have only limited bearing on
phenomenology, but we lay out here the potential for misinterpretation
in this channel and look forward to a relevant discussion of these
matters in a future analysis with improved systematics. We note that an
analysis varying the lattice spacing using the same operator set might
shed light on the spin-assignment - no operators are used with have
overlap on to spin-4 in the continuum, any overlap with such a
state is suppressed by powers of $a$ that will reduce as $a \to
0$. Thus a rapidly reducing $Z(a)$ would suggest a spin-4
interpretation.


\begin{acknowledgments}
We thank Eric Swanson for his results solving a charmonium potential
model in a finite box.

Notice: Authored by Jefferson Science Associates, LLC under U.S. DOE
Contract No. DE-AC05-06OR23177. The U.S. Government retains a
non-exclusive, paid-up, irrevocable, world-wide license to publish or
reproduce this manuscript for U.S. Government purposes. Computations were performed
on clusters at Jefferson Laboratory as part of the USQCD
collaboration.

\end{acknowledgments}

\appendix

\section{Continuum overlaps}\label{overlaps}

In this appendix we tabulate the Lorentz covariant kinematic
structures corresponding to overlap of fermion bilinear interpolating fields with
a limited number of Lorentz indices and states of definite
$J^{PC}$. In all cases the fermion fields should be considered to be
evaluated at the origin of Minkowski space time.

\subsection{No Lorentz Indices}
\begin{align}
\langle 0 | \bar{\psi} \psi | 0^{++}(\vec{p}) \rangle &= Z \nonumber \\
\langle 0 | \bar{\psi} \gamma^5 \psi | 0^{-+}(\vec{p}) \rangle &= Z  \nonumber
\end{align} 

\subsection{One Lorentz index}
\begin{align}
\langle 0 | \bar{\psi} {\cal O}^\mu \psi | 0^{\mathbb{PC}}(\vec{p})
\rangle &= Z p^\mu  \nonumber\\
\langle 0 | \bar{\psi} {\cal O}^\mu \psi | 1^{\mathbb{(-P)C}}(\vec{p},
r) \rangle &= Z \in^\mu \hspace{-1mm}(\vec{p},r) \nonumber
\end{align} 
\begin{tabular}{c|cccc}
${\cal O}^\mu $ & $\gamma^\mu$ & $\overleftrightarrow{D}^\mu$ &  $\gamma^5 \gamma^\mu$ & $\gamma^5 \overleftrightarrow{D}^\mu$  \\
\hline
$\mathbb{P}\mathbb{C}$ & $+-$  & $+-$ & $-+$ & $--$ 
\end{tabular}

\subsection{Two Lorentz indices}
\begin{align}
\langle 0 | \bar{\psi} {\cal O}^{\mu\nu} \psi | 0^{\mathbb{PC}}(\vec{p})
\rangle &= Z_0 g^{\mu\nu} + Z_p p^\mu p^\nu  \nonumber \\
\langle 0 | \bar{\psi} {\cal O}^{\mu\nu} \psi | 1^{\mathbb{PC}}(\vec{p},
r) \rangle &= Z \epsilon^{\mu\nu\alpha\beta} \in_\alpha\hspace{-1mm}(\vec{p},r)
p_\beta  \nonumber\\
\langle 0 | \bar{\psi} {\cal O}^{\mu\nu} \psi | 2^{\mathbb{PC}}(\vec{p},
r) \rangle &= Z \in^{\mu\nu}(\vec{p},r)  \nonumber\\
\langle 0 | \bar{\psi} {\cal O}^{\mu\nu} \psi | 1^{\mathbb{(-P)C}}(\vec{p},
r) \rangle &= Z_+ \big( \in^\mu\hspace{-1mm}(\vec{p},r) p^\nu +  \in^\nu\hspace{-1mm}(\vec{p},r)
p^\mu \big) + Z_- \big( \in^\mu\hspace{-1mm}(\vec{p},r) p^\nu -  \in^\nu\hspace{-1mm}(\vec{p},r)
p^\mu \big) \nonumber
\end{align} 
\begin{tabular}{c|ccccccc}
${\cal O}^{\mu\nu}$ & $[\gamma^\mu, \gamma^\nu]$ & $\gamma^\mu \overleftrightarrow{D}^\nu$ & $\gamma^5 \gamma^\mu \overleftrightarrow{D}^\nu$ & $\{ \overleftrightarrow{D}^\mu , \overleftrightarrow{D}^\nu \}$ & $\gamma^5 \{ \overleftrightarrow{D}^\mu , \overleftrightarrow{D}^\nu \}$  & $\gamma^5 [ \overleftrightarrow{D}^\mu , \overleftrightarrow{D}^\nu ]$  & $\gamma^5 [ \overleftrightarrow{D}^\mu , \overleftrightarrow{D}^\nu ]$  \\
\hline
 $\mathbb{P}\mathbb{C}$& $+-$  & $++$  & $--$  & $++$& $-+$& $+-$& $--$
\end{tabular}

\subsection{Three Lorentz indices}
\begin{align}
\langle 0 | \bar{\psi} {\cal O}^{\mu\nu\rho} \psi | 0^{\mathbb{PC}}(\vec{p})
\rangle &= Z_0 g^{\mu\nu} p^\rho + Z_+ \big( g^{\mu\rho} p^\nu +
g^{\nu\rho} p^\mu \big) + Z_- \big( g^{\mu\rho} p^\nu -
g^{\nu\rho} p^\mu \big) + Z_p p^\mu p^\nu p^\rho  \nonumber \\
\langle 0 | \bar{\psi} {\cal O}^{\mu\nu\rho} \psi | 0^{\mathbb{(-P)C}}(\vec{p})
\rangle &= Z \epsilon^{\mu\nu\rho\alpha} p_\alpha   \nonumber\\
\langle 0 | \bar{\psi} {\cal O}^{\mu\nu\rho} \psi | 1^{\mathbb{PC}}(\vec{p},r)
\rangle &= Z_0 \epsilon^{\mu\nu\rho\alpha} \in_\alpha\hspace{-1mm}(\vec{p},r)   +
Z_{p0} \epsilon^{\mu\nu\alpha\beta} p_\alpha \in_\beta\hspace{-1mm}(\vec{p},r)
p^\rho  \nonumber\\
&+ Z_+ \big(\epsilon^{\mu\rho\alpha\beta} p_\alpha \in_\beta\hspace{-1mm}(\vec{p},r)
p^\nu + \epsilon^{\nu\rho\alpha\beta} p_\alpha \in_\beta\hspace{-1mm}(\vec{p},r)
p^\mu \big)   + Z_- \big(\epsilon^{\mu\rho\alpha\beta} p_\alpha \in_\beta\hspace{-1mm}(\vec{p},r)
p^\nu - \epsilon^{\nu\rho\alpha\beta} p_\alpha \in_\beta\hspace{-1mm}(\vec{p},r)
p^\mu \big)  \nonumber\\
\langle 0 | \bar{\psi} {\cal O}^{\mu\nu\rho} \psi | 1^{\mathbb{(-P)C}}(\vec{p},r)
\rangle &= Z_0 g^{\mu\nu} \in^\rho\hspace{-1mm}(\vec{p},r) +  Z_+ \big( g^{\mu\rho} \in^\nu\hspace{-1mm}(\vec{p},r) +
g^{\nu\rho} \in^\mu\hspace{-1mm}(\vec{p},r) \big) + Z_- \big( g^{\mu\rho} \in^\nu\hspace{-1mm}(\vec{p},r) -
g^{\nu\rho} \in^\mu\hspace{-1mm}(\vec{p},r) \big)    \nonumber\\
&+ Z_{p0} p^\mu p^\nu
\in^\rho\hspace{-1mm}(\vec{p},r) +  Z_{p+} \big( p^\mu p^\rho \in^\nu\hspace{-1mm}(\vec{p},r) +
p^\nu p^\rho \in^\mu\hspace{-1mm}(\vec{p},r) \big) + Z_{p-} \big( p^\mu p^\rho \in^\nu\hspace{-1mm}(\vec{p},r) -
p^\nu p^\rho \in^\mu\hspace{-1mm}(\vec{p},r) \big)   \nonumber\\
\langle 0 | \bar{\psi} {\cal O}^{\mu\nu\rho} \psi | 2^{\mathbb{PC}}(\vec{p},r)
\rangle &= Z_0 \in^{\mu\nu}\hspace{-1mm}(\vec{p},r) p^\rho + Z_+ \big(
\in^{\mu\rho}\hspace{-1mm}(\vec{p},r) p^\nu + \in^{\nu\rho}\hspace{-1mm}(\vec{p},r) p^\mu \big)
+ Z_- \big(
\in^{\mu\rho}\hspace{-1mm}(\vec{p},r) p^\nu - \in^{\nu\rho}\hspace{-1mm}(\vec{p},r) p^\mu
\big) \nonumber\\
\langle 0 | \bar{\psi} {\cal O}^{\mu\nu\rho} \psi | 2^{\mathbb{(-P)C}}(\vec{p},r)
\rangle &= Z_0 \epsilon^{\mu\nu\alpha\beta} \in_\alpha^\rho\hspace{-1mm}(\vec{p},r)
p_\beta + Z_+ \big(    \epsilon^{\mu\rho\alpha\beta} \in_\alpha^\nu\hspace{-1mm}(\vec{p},r)
p_\beta  + \epsilon^{\nu\rho\alpha\beta} \in_\alpha^\mu\hspace{-1mm}(\vec{p},r)
p_\beta \big)   \nonumber\\
&+ Z_- \big(    \epsilon^{\mu\rho\alpha\beta} \in_\alpha^\nu\hspace{-1mm}(\vec{p},r)
p_\beta  - \epsilon^{\nu\rho\alpha\beta} \in_\alpha^\mu\hspace{-1mm}(\vec{p},r)
p_\beta \big) \nonumber\\
\langle 0 | \bar{\psi} {\cal O}^{\mu\nu\rho} \psi | 3^{\mathbb{(-P)C}}(\vec{p},r)
\rangle &= Z \in^{\mu\nu\rho}\hspace{-1mm}(\vec{p},r) \nonumber
\end{align} 
\begin{tabular}{c|ccccc}
${\cal O}^{\mu\nu\rho}$ &$\gamma^\mu \gamma^\nu \overleftrightarrow{D}^\rho$  & $\{ \overleftrightarrow{D}^\mu ,  \overleftrightarrow{D}^\nu\}
\gamma^\rho$ & $\{ \overleftrightarrow{D}^\mu ,  \overleftrightarrow{D}^\nu\}
\gamma^5 \gamma^\rho$ & $[\overleftrightarrow{D}^\mu ,  \overleftrightarrow{D}^\nu]
\gamma^\rho$ & $[ \overleftrightarrow{D}^\mu ,  \overleftrightarrow{D}^\nu ]
\gamma^5 \gamma^\rho$ \\
\hline
 $\mathbb{P}\mathbb{C}$ & $++$& $+-$ & $-+$ & $++$ & $--$
\end{tabular}

\subsection{Four Lorentz indices}
Here we limit ourselves to the case where the first two indices are
antisymmetric and the final two have definite symmetry.

\begin{align}
\langle 0 | \bar{\psi} {\cal O}^{\mu\nu\rho\sigma} \psi | 0^{\mathbb{PC}}(\vec{p})
\rangle &= Z_0 \big( g^{\mu\rho} g^{\nu\sigma} -  g^{\nu\rho}
g^{\mu\sigma} \big) + Z_{p\pm} \Big[   \big(   g^{\mu\rho}
p^\nu p^\sigma -    g^{\nu\rho} p^\mu p^\sigma  \big)   \pm  \big(   g^{\mu\sigma}
p^\nu p^\rho -    g^{\nu\sigma} p^\mu p^\rho  \big)    \Big]  \nonumber\\
\langle 0 | \bar{\psi} {\cal O}^{\mu\nu\rho\sigma} \psi | 0^{\mathbb{(-P)C}}(\vec{p})
\rangle &= Z_0 \epsilon^{\mu\nu\rho\sigma} + Z_p \big(
\epsilon^{\mu\rho \sigma \alpha} p_\alpha p^\nu -  \epsilon^{\nu\rho
  \sigma \alpha} p_\alpha p^\mu   \big) + Z_\pm \big(
\epsilon^{\mu\nu \rho\alpha} p_\alpha p^\sigma \pm \epsilon^{\mu\nu
  \sigma\alpha} p_\alpha p^\rho   \big)  \nonumber\\
\langle 0 | \bar{\psi} {\cal O}^{\mu\nu\rho\sigma} \psi | 1^{\mathbb{PC}}(\vec{p},r)
\rangle &= Z_{p\pm} \big(  \epsilon^{\mu\nu\rho\alpha}
\in_\alpha\hspace{-1mm}(\vec{p},r) p^\sigma \pm \epsilon^{\mu\nu\sigma\alpha}
\in_\alpha\hspace{-1mm}(\vec{p},r) p^\rho \big) + Z_{\in \pm} \big(  \epsilon^{\mu\nu\rho\alpha}
 p_\alpha \in^\sigma \hspace{-1mm}(\vec{p},r) \pm \epsilon^{\mu\nu\sigma\alpha}
p_\alpha \in^\rho\hspace{-1mm}(\vec{p},r)  \big)  \nonumber\\
&+ Z_p \big(  \epsilon^{\rho\sigma\mu\alpha}
\in_\alpha\hspace{-1mm}(\vec{p},r) p^\nu \pm \epsilon^{\rho\sigma\nu\alpha}
\in_\alpha\hspace{-1mm}(\vec{p},r) p^\mu \big) + Z_{\in} \big(  \epsilon^{\rho\sigma\mu\alpha}
 p_\alpha \in^\nu \hspace{-1mm}(\vec{p},r) \pm \epsilon^{\rho\sigma\nu\alpha}
p_\alpha \in^\mu\hspace{-1mm}(\vec{p},r)  \big)  \nonumber\\
&+ Z_g g^{\rho\sigma} \epsilon^{\mu\nu\alpha\beta}
\in_\alpha\hspace{-1mm}(\vec{p},r) p_\beta + Z_{p^2} p^\rho p^\sigma \epsilon^{\mu\nu\alpha\beta}
\in_\alpha\hspace{-1mm}(\vec{p},r) p_\beta  \nonumber\\
&\hspace{-15mm}+ Z_{gp\pm} \Big[  \big(   g^{\mu\rho}
\epsilon^{\nu\sigma\alpha\beta}  \in_\alpha\hspace{-1mm}(\vec{p},r)
p_\beta  -  g^{\nu\rho}
\epsilon^{\mu\sigma\alpha\beta}  \in_\alpha\hspace{-1mm}(\vec{p},r)
p_\beta \big)   \pm    \big(   g^{\mu\sigma}
\epsilon^{\nu\rho\alpha\beta}  \in_\alpha\hspace{-1mm}(\vec{p},r)
p_\beta  -  g^{\nu\sigma}
\epsilon^{\mu\rho\alpha\beta}  \in_\alpha\hspace{-1mm}(\vec{p},r)
p_\beta \big)    \Big]  \nonumber\\
&\hspace{-15mm}+  Z_{p^2\pm} \Big[  \big(   p^\mu p^\rho
\epsilon^{\nu\sigma\alpha\beta}  \in_\alpha\hspace{-1mm}(\vec{p},r)
p_\beta  -  p^\nu p^\rho
\epsilon^{\mu\sigma\alpha\beta}  \in_\alpha\hspace{-1mm}(\vec{p},r)
p_\beta \big)   \pm    \big(   p^\mu p^\sigma
\epsilon^{\nu\rho\alpha\beta}  \in_\alpha\hspace{-1mm}(\vec{p},r)
p_\beta  -  p^\nu p^\sigma
\epsilon^{\mu\rho\alpha\beta}  \in_\alpha\hspace{-1mm}(\vec{p},r)
p_\beta \big)    \Big]  \nonumber\\
\langle 0 | \bar{\psi} {\cal O}^{\mu\nu\rho\sigma} \psi | 1^{\mathbb{(-P)C}}(\vec{p},r)
\rangle &= Z_g  g^{\rho\sigma} \big(  \in^\mu\hspace{-1mm}(\vec{p},r)
p^\nu -  \in^\nu\hspace{-1mm}(\vec{p},r)
p^\mu  \big)    + Z_{p^2}   p^\rho p^\sigma \big(  \in^\mu\hspace{-1mm}(\vec{p},r)
p^\nu -  \in^\nu\hspace{-1mm}(\vec{p},r)
p^\mu  \big)    \nonumber\\
&+ Z_{1\pm}  \Big[  \big( g^{\mu\rho} \in^\nu\hspace{-1mm}(\vec{p},r)
p^\sigma -  g^{\nu\rho} \in^\mu\hspace{-1mm}(\vec{p},r)
p^\sigma  \big)  \pm   \big( g^{\mu\sigma} \in^\nu\hspace{-1mm}(\vec{p},r)
p^\rho -  g^{\nu\sigma} \in^\mu\hspace{-1mm}(\vec{p},r)
p^\rho  \big)   \Big]  \nonumber\\
&+ Z_{2\pm}  \Big[  \big(  g^{\nu\sigma}
\in^\rho\hspace{-1mm}(\vec{p},r) p^\mu -  g^{\mu\sigma}
\in^\rho\hspace{-1mm}(\vec{p},r) p^\nu \big)  \pm  \big(  g^{\nu\rho}
\in^\sigma\hspace{-1mm}(\vec{p},r) p^\mu -  g^{\mu\rho}
\in^\sigma\hspace{-1mm}(\vec{p},r) p^\nu \big)   \Big]  \nonumber\\
\langle 0 | \bar{\psi} {\cal O}^{\mu\nu\rho\sigma} \psi | 2^{\mathbb{PC}}(\vec{p},r)
\rangle &= Z_{g\pm} \Big[ \big(  g^{\mu\rho}
\in^{\nu\sigma}\hspace{-1mm}(\vec{p},r) - g^{\nu\rho}
\in^{\mu\sigma}\hspace{-1mm}(\vec{p},r)    \big)    \pm \big(  g^{\mu\sigma}
\in^{\nu\rho}\hspace{-1mm}(\vec{p},r) - g^{\nu\sigma}
\in^{\mu\rho}\hspace{-1mm}(\vec{p},r)    \big)  \Big]  \nonumber\\
&+ Z_{p^2\pm} \Big[ \big(  p^\mu p^\rho
\in^{\nu\sigma}\hspace{-1mm}(\vec{p},r) - p^\nu p^\rho
\in^{\mu\sigma}\hspace{-1mm}(\vec{p},r)    \big)    \pm \big(  p^\mu p^\sigma
\in^{\nu\rho}\hspace{-1mm}(\vec{p},r) - p^\nu p^\sigma
\in^{\mu\rho}\hspace{-1mm}(\vec{p},r)    \big)  \Big]  \nonumber\\
\langle 0 | \bar{\psi} {\cal O}^{\mu\nu\rho\sigma} \psi | 2^{\mathbb{(-P)C}}(\vec{p},r)
\rangle &= Z_{\pm} \big(  \epsilon^{\mu\nu\rho\alpha}
\in_\alpha^\sigma\hspace{-1mm}(\vec{p},r)  \pm \epsilon^{\mu\nu\sigma\alpha}
\in_\alpha^\rho\hspace{-1mm}(\vec{p},r)    \big) + Z_0 \big(  \epsilon^{\mu\rho\sigma\alpha}
\in_\alpha^\nu\hspace{-1mm}(\vec{p},r) - \epsilon^{\nu\rho\sigma\alpha}
\in_\alpha^\mu\hspace{-1mm}(\vec{p},r) \big)  \nonumber\\
&+ Z_{1\pm} \big(  \epsilon^{\mu\nu\alpha\beta}
p_\alpha \in_\beta^\rho\hspace{-1mm}(\vec{p},r) p^\sigma \pm  \epsilon^{\mu\nu\alpha\beta}
p_\alpha \in_\beta^\sigma\hspace{-1mm}(\vec{p},r) p^\rho  \nonumber\\
&+  Z_{2\pm} \Big[   \big(  \epsilon^{\mu\rho\alpha\beta} p_\alpha
\in_\beta^\nu\hspace{-1mm}(\vec{p},r)   p^\sigma - \epsilon^{\nu\rho\alpha\beta} p_\alpha
\in_\beta^\mu\hspace{-1mm}(\vec{p},r)   p^\sigma\big)  \pm  \big(  \epsilon^{\mu\sigma\alpha\beta} p_\alpha
\in_\beta^\nu\hspace{-1mm}(\vec{p},r)   p^\rho - \epsilon^{\nu\sigma\alpha\beta} p_\alpha
\in_\beta^\mu\hspace{-1mm}(\vec{p},r)   p^\rho\big)  \Big] \nonumber\\
\langle 0 | \bar{\psi} {\cal O}^{\mu\nu\rho\sigma} \psi | 3^{\mathbb{PC}}(\vec{p},r)
\rangle &= Z_0 \epsilon^{\mu\nu\alpha\beta} p_\alpha
\in_\beta^{\rho\sigma}\hspace{-1mm}(\vec{p},r)  \nonumber\\
&+ Z_\pm \Big[ \big(  \epsilon^{\mu\rho\alpha\beta} p_\alpha
\in_\beta^{\nu\sigma}\hspace{-1mm}(\vec{p},r)  -
\epsilon^{\nu\rho\alpha\beta} p_\alpha
\in_\beta^{\mu\sigma}\hspace{-1mm}(\vec{p},r) \big)   \pm \big(  \epsilon^{\mu\sigma\alpha\beta} p_\alpha
\in_\beta^{\nu\rho}\hspace{-1mm}(\vec{p},r)  -
\epsilon^{\nu\sigma\alpha\beta} p_\alpha
\in_\beta^{\mu\rho}\hspace{-1mm}(\vec{p},r) \big)  \Big]  \nonumber\\
\langle 0 | \bar{\psi} {\cal O}^{\mu\nu\rho\sigma} \psi | 3^{\mathbb{(-P)C}}(\vec{p},r)
\rangle &= Z_0 \epsilon^{\mu\nu\alpha\beta} p_\alpha\big(  p^\mu
\in^{\nu\rho\sigma}\hspace{-1mm}(\vec{p},r)  -  p^\nu
\in^{\mu\rho\sigma}\hspace{-1mm}(\vec{p},r)   \big) \nonumber\\
\langle 0 | \bar{\psi} {\cal O}^{\mu\nu\rho\sigma} \psi | 4^{\mathbb{PC}}(\vec{p},r)
\rangle &= 0 \nonumber
\end{align}
\begin{tabular}{c|cc}
${\cal O}^{\mu\nu\rho\sigma}$ & $\gamma^\mu \gamma^\nu [\overleftrightarrow{D}^\rho, \overleftrightarrow{D}^\sigma]$ & $\gamma^\mu \gamma^\nu \{\overleftrightarrow{D}^\rho, \overleftrightarrow{D}^\sigma\}$ \\
\hline
$\mathbb{P}\mathbb{C}$ & $+-$ & $++$ 
\end{tabular}

\subsection{Minkowski hermitian operators}
The following operators are hermitian in Minkowski space. Using these
operators one is guaranteed a hermitian correlator matrix which is a
requirement for the variational method.

\begin{tabular}{ccccc}
$\bar{\psi}\psi$ & $\bar{\psi}i \gamma^5 \psi$ & $\bar{\psi}\gamma^\mu
\psi$ & $\bar{\psi}\gamma^5 \gamma^\mu\psi$ & $\bar{\psi} i \gamma^\mu \gamma^\nu\psi$.
\end{tabular}

\begin{tabular}{ccccc}
$\bar{\psi}i \overleftrightarrow{D_i}\psi$ & $\bar{\psi}
\gamma^5\overleftrightarrow{D_i} \psi$ & $\bar{\psi}i \gamma^\mu\overleftrightarrow{D_i}
\psi$ & $\bar{\psi}i \gamma^5 \gamma^\mu\overleftrightarrow{D_i}\psi$ & $\bar{\psi} \gamma^\mu \gamma^\nu\overleftrightarrow{D_i}\psi$.
\end{tabular}

\begin{tabular}{ccccc}
$\bar{\psi} \overleftrightarrow{\mathbb{D}_i}\psi$ & $\bar{\psi}i
\gamma^5  \overleftrightarrow{\mathbb{D}_i}\psi$ &
$\bar{\psi}\gamma^\mu  \overleftrightarrow{\mathbb{D}_i}
\psi$ & $\bar{\psi}\gamma^5 \gamma^\mu
\overleftrightarrow{\mathbb{D}_i} \psi$ & $\bar{\psi} i \gamma^\mu
\gamma^\nu  \overleftrightarrow{\mathbb{D}_i}\psi$.
\end{tabular}

\begin{tabular}{ccccc}
$\bar{\psi} \overleftrightarrow{\mathbb{E}_i}\psi$ & $\bar{\psi}i
\gamma^5  \overleftrightarrow{\mathbb{E}_i}\psi$ &
$\bar{\psi}\gamma^\mu  \overleftrightarrow{\mathbb{E}_i}
\psi$ & $\bar{\psi}\gamma^5 \gamma^\mu
\overleftrightarrow{\mathbb{E}_i} \psi$ & $\bar{\psi} i \gamma^\mu
\gamma^\nu  \overleftrightarrow{\mathbb{E}_i}\psi$.
\end{tabular}

\begin{tabular}{ccccc}
$\bar{\psi}i \mathbb{B}_i\psi$ & $\bar{\psi}
\gamma^5\mathbb{B}_i\psi$ & $\bar{\psi}i \gamma^\mu\mathbb{B}_i
\psi$ & $\bar{\psi}i \gamma^5 \gamma^\mu\mathbb{B}_i\psi$ & $\bar{\psi} \gamma^\mu \gamma^\nu\mathbb{B}_i\psi$.
\end{tabular}

\section{Lattice irrep operators at $\vec{p}=(000)$}\label{ops}

Third column indicates the quantum numbers accessible in the continuum limit.

\begin{tabular}{ccc|ccc|ccc}
op. & name & cont.  & op. & name & cont. & op. & name & cont.\\
\hline
$\nabla^i$ & $(a_0 \times \nabla)_{T_1}$ & $1^{--}$ & $\mathbb{D}^i $ & $(a_0
\times \mathbb{D})_{T_2}$ & $2^{++}$ & $\mathbb{B}^i$ & $(a_0
\times \mathbb{B})_{T_1}$ & $1^{+-}$ \\
$\gamma^5 \nabla^i$ & $(\pi \times \nabla)_{T_1}$ & $1^{+-}$  &
$\gamma^5 \mathbb{D}^i $ & $(\pi
\times \mathbb{D})_{T_2}$ & $2^{-+}$ & $\gamma^5 \mathbb{B}^i$ & $(\pi
\times \mathbb{B})_{T_1}$ & $1^{--}$ \\
$\gamma^4 \gamma^5 \nabla^i$ & $(\pi_{(2)} \times \nabla)_{T_1}$ &
$1^{+-}$ & $\gamma^4 \gamma^5 \mathbb{D}^i $ & $(\pi_{(2)}
\times \mathbb{D})_{T_2}$ & $2^{-+}$ & $\gamma^4 \gamma^5 \mathbb{B}^i$ & $(\pi_{(2)}
\times \mathbb{B})_{T_1}$ & $1^{--}$ \\
$\gamma^4 \nabla^i$ & $(a_{0(2)} \times \nabla)_{T_1}$ & $1^{-+}$ &
$\gamma^4 \mathbb{D}^i $ & $(a_{0(2)}
\times \mathbb{D})_{T_2}$ & $2^{+-}$ & $\gamma^4 \mathbb{B}^i$ & $(a_{0(2)}
\times \mathbb{B})_{T_1}$ & $1^{++}$ \\
&&&&&&&&\\
$\gamma^i \nabla^i$ & $(\rho \times \nabla)_{A_1}$ & $0^{++}$    & $\gamma^i \mathbb{D}^i$ & $(\rho \times \mathbb{D})_{A_2}$ & $3^{--}$ & $\gamma^i \mathbb{B}^i$ & $(\rho \times \mathbb{B})_{A_1}$ & $0^{-+}$          \\           
$\epsilon_{ijk} \gamma^j \nabla^k$ & $(\rho \times \nabla)_{T_1}$   & $1^{++}$  &  $|\epsilon_{ijk}| \gamma^j \mathbb{D}^k$ & $(\rho \times \mathbb{D})_{T_1}$ & $1^{--}$ &  $\epsilon_{ijk} \gamma^j \mathbb{B}^k$ & $(\rho \times \mathbb{B})_{T_1}$ & $1^{-+}$\\       
$|\epsilon_{ijk}| \gamma^j \nabla^k$ & $(\rho \times \nabla)_{T_2}$ & $2^{++}$ & $\epsilon_{ijk} \gamma^j \mathbb{D}^k$ & $(\rho \times \mathbb{D})_{T_2}$ & $(2,3)^{--}$   & $|\epsilon_{ijk}| \gamma^j \mathbb{B}^k$ & $(\rho \times \mathbb{B})_{T_2}$ & $2^{-+}$\\ 
$\mathbb{Q}_{ijk} \gamma^j \nabla^k$ & $(\rho \times \nabla)_{E}$ & $2^{++}$ &  $\mathbb{Q}_{ijk} \gamma^j \mathbb{D}^k$ & $(\rho \times \mathbb{D})_{E}$ & $2^{--}$   &  $\mathbb{Q}_{ijk} \gamma^j \mathbb{B}^k$ & $(\rho \times \mathbb{B})_{E}$ & $2^{-+}$\\       
&&&&&&&&\\
$\gamma^4 \gamma^i \nabla^i$ & $(\rho_{(2)} \times \nabla)_{A_1}$ & $0^{++}$                          & $\gamma^4\gamma^i \mathbb{D}^i$ & $(\rho_{(2)} \times \mathbb{D})_{A_2}$ & $3^{--}$ & $\gamma^4 \gamma^i \mathbb{B}^i$ & $(\rho_{(2)} \times \mathbb{B})_{A_1}$ & $0^{-+}$          \\                                        
$\epsilon_{ijk} \gamma^4 \gamma^j \nabla^k$ & $(\rho_{(2)} \times \nabla)_{T_1}$ & $1^{++}$      &  $|\epsilon_{ijk}|\gamma^4 \gamma^j \mathbb{D}^k$ & $(\rho_{(2)} \times \mathbb{D})_{T_1}$ & $1^{--}$ &  $\epsilon_{ijk} \gamma^4\gamma^j \mathbb{B}^k$ & $(\rho_{(2)} \times \mathbb{B})_{T_1}$ & $1^{-+}$\\   
$|\epsilon_{ijk}| \gamma^4 \gamma^j \nabla^k$ & $(\rho_{(2)} \times \nabla)_{T_2}$ & $2^{++}$     & $\epsilon_{ijk} \gamma^4 \gamma^j \mathbb{D}^k$ & $(\rho_{(2)} \times \mathbb{D})_{T_2}$ & $(2,3)^{--}$   & $|\epsilon_{ijk}| \gamma^4\gamma^j \mathbb{B}^k$ & $(\rho_{(2)} \times \mathbb{B})_{T_2}$ & $2^{-+}$\\
$\mathbb{Q}_{ijk} \gamma^4 \gamma^j \nabla^k$ & $(\rho_{(2)} \times \nabla)_{E}$ &$2^{++}$        &  $\mathbb{Q}_{ijk} \gamma^4 \gamma^j \mathbb{D}^k$ & $(\rho_{(2)} \times \mathbb{D})_{E}$ & $2^{--}$   &  $\mathbb{Q}_{ijk} \gamma^4\gamma^j \mathbb{B}^k$ & $(\rho_{(2)} \times \mathbb{B})_{E}$ & $2^{-+}$\\    
&&&&&&&&\\
$\gamma^5 \gamma^i \nabla^i$ & $(a_1 \times \nabla)_{A_1}$ & $0^{--}$                &       $\gamma^5 \gamma^i \mathbb{D}^i$ & $(a_1 \times \mathbb{D})_{A_2}$ & $3^{++}$    &       $\gamma^5 \gamma^i \mathbb{B}^i$ & $(a_1 \times \mathbb{B})_{A_1}$ & $0^{+-}$                           \\                   
$\epsilon_{ijk} \gamma^5 \gamma^j \nabla^k$ & $(a_1 \times \nabla)_{T_1}$ & $1^{--}$    &       $|\epsilon_{ijk}| \gamma^5 \gamma^j \mathbb{D}^k$ & $(a_1 \times \mathbb{D})_{T_1}$ & $1^{++}$      &             $\epsilon_{ijk} \gamma^5 \gamma^j \mathbb{B}^k$ & $(a_1 \times \mathbb{B})_{T_1}$ & $1^{+-}$    \\                  
$|\epsilon_{ijk}| \gamma^5 \gamma^j \nabla^k$ & $(a_1 \times \nabla)_{T_2}$ & $2^{--}$ &     $\epsilon_{ijk} \gamma^5 \gamma^j \mathbb{D}^k$ & $(a_1 \times \mathbb{D})_{T_2}$ & $(2,3)^{++}$      &          $|\epsilon_{ijk}| \gamma^5 \gamma^j \mathbb{B}^k$ & $(a_1 \times \mathbb{B})_{T_2}$ & $2^{+-}$    \\               
$\mathbb{Q}_{ijk} \gamma^5 \gamma^j \nabla^k$ & $(a_1 \times \nabla)_{E}$ & $2^{--}$   &     $\mathbb{Q}_{ijk} \gamma^5 \gamma^j \mathbb{D}^k$ & $(a_1 \times \mathbb{D})_{E}$ & $2^{++}$       &             $\mathbb{Q}_{ijk} \gamma^5 \gamma^j \mathbb{B}^k$ & $(a_1 \times \mathbb{B})_{E}$ & $2^{+-}$    \\                        
&&&&&&&&\\
$\gamma^4 \gamma^5 \gamma^i \nabla^i$ & $(b_1 \times \nabla)_{A_1}$ & $0^{-+}$                         &       $\gamma^4 \gamma^5 \gamma^i \mathbb{D}^i$ & $(b_1 \times \mathbb{D})_{A_2}$ & $3^{+-}$                         &                 $\gamma^4 \gamma^5 \gamma^i \mathbb{B}^i$ & $(b_1 \times \mathbb{B})_{A_1}$ & $0^{++}$          \\              
$\epsilon_{ijk} \gamma^4\gamma^5 \gamma^j \nabla^k$ & $(b_1 \times \nabla)_{T_1}$ & $1^{-+}$      &      $|\epsilon_{ijk}| \gamma^4\gamma^5 \gamma^j \mathbb{D}^k$ & $(b_1 \times \mathbb{D})_{T_1}$ & $1^{+-}$      &                             $\epsilon_{ijk} \gamma^4\gamma^5 \gamma^j \mathbb{B}^k$ & $(b_1 \times \mathbb{B})_{T_1}$ & $1^{++}$      \\        
$|\epsilon_{ijk}| \gamma^4\gamma^5 \gamma^j \nabla^k$ & $(b_1 \times \nabla)_{T_2}$ & $2^{-+}$     &    $\epsilon_{ijk} \gamma^4\gamma^5 \gamma^j \mathbb{D}^k$ & $(b_1 \times \mathbb{D})_{T_2}$ & $(2,3)^{+-}$     &                        $|\epsilon_{ijk}| \gamma^4\gamma^5 \gamma^j \mathbb{B}^k$ & $(b_1 \times \mathbb{B})_{T_2}$ & $2^{++}$     \\     
$\mathbb{Q}_{ijk} \gamma^4\gamma^5 \gamma^j \nabla^k$ & $(b_1 \times \nabla)_{E}$ & $2^{-+}$    &      $\mathbb{Q}_{ijk} \gamma^4\gamma^5 \gamma^j \mathbb{D}^k$ & $(b_1 \times \mathbb{D})_{E}$ & $2^{+-}$    &                               $\mathbb{Q}_{ijk} \gamma^4\gamma^5 \gamma^j \mathbb{B}^k$ & $(b_1 \times \mathbb{B})_{E}$ & $2^{++}$    \\             
\end{tabular}

\begin{tabular}{ccc}
op. & name & cont. \\
\hline
$\mathbb{E}^i$ & $(a_0 \times \mathbb{E})_{T_2}$ & $2^{++}$\\
$\gamma^5 \mathbb{E}^i$ & $(\pi \times \mathbb{E})_{T_2}$ & $2^{-+}$\\
$\gamma^4 \gamma^5 \mathbb{E}^i$ & $(\pi_{(2)} \times \mathbb{E})_{T_2}$ & $2^{-+}$\\
$\gamma^4 \mathbb{E}^i$ & $(a_{0(2)} \times \mathbb{E})_{T_2}$ & $2^{+-}$\\
\\
$\mathbb{R}_{ijk} \gamma^j \mathbb{E}^k$ & $(\rho \times \mathbb{E})_{T_1}$ & $(1,3)^{--}$\\
$\mathbb{T}_{ijk} \gamma^j \mathbb{E}^k$ & $(\rho \times \mathbb{E})_{T_2}$ & $(2,3)^{--}$\\
\\
$\mathbb{R}_{ijk} \gamma^4 \gamma^j \mathbb{E}^k$ & $(\rho_{(2)} \times \mathbb{E})_{T_1}$ & $(1,3)^{--}$\\
$\mathbb{T}_{ijk} \gamma^4 \gamma^j \mathbb{E}^k$ & $(\rho_{(2)} \times \mathbb{E})_{T_2}$ & $(2,3)^{--}$\\
\\
$\mathbb{R}_{ijk} \gamma^5 \gamma^j \mathbb{E}^k$ & $(a_1 \times \mathbb{E})_{T_1}$ & $(1,3)^{++}$\\
$\mathbb{T}_{ijk} \gamma^5 \gamma^j \mathbb{E}^k$ & $(a_1 \times \mathbb{E})_{T_2}$ & $(2,3)^{++}$\\
\\
$\mathbb{R}_{ijk} \gamma^4 \gamma^5 \gamma^j \mathbb{E}^k$ & $(b_1 \times \mathbb{E})_{T_1}$ & $(1,3)^{+-}$\\
$\mathbb{T}_{ijk} \gamma^4 \gamma^5 \gamma^j \mathbb{E}^k$ & $(b_1 \times \mathbb{E})_{T_2}$ & $(2,3)^{+-}$\\
\end{tabular}

\vspace{.7cm}
$\mathbb{R}_{ijk}, \mathbb{T}_{ijk}$ are the Clebsch-Gordan
coefficients for $T_1 \otimes E = T_1 \oplus T_2$. We find the non-zero
values
\begin{multline}
\mathbb{R}_{111} = \tfrac{\sqrt{3}}{4};\;\; \mathbb{R}_{112} = -
  \tfrac{1}{4};\;\; \mathbb{R}_{221} = -\tfrac{\sqrt{3}}{4};\;\; \mathbb{R}_{222} = -\tfrac{1}{4};\;\;  \mathbb{R}_{332} = \tfrac{1}{2},\\
\mathbb{T}_{111} = \tfrac{1}{4};\;\; \mathbb{T}_{112} = 
  \tfrac{\sqrt{3}}{4};\;\; \mathbb{T}_{221} = \tfrac{1}{4};\;\; \mathbb{T}_{222} = -\tfrac{\sqrt{3}}{4};\;\;  \mathbb{T}_{332} = -\tfrac{1}{2}.
\end{multline}

\section{Variational analysis of toy data}\label{toy}
In this section we consider the variational method applied to a finite
spectrum of ten states when using a set of only five operators.

We use the mass spectrum $m_\alpha=$(0.5, 0.6, 0.63, 0.7, 0.72, 0.8,  0.83, 0.9,
0.91, 1.04) proposing that there exist ideal operators which have
unit-normalized overlap on to only one state, i.e. there are
$\tilde{Z}^\alpha_i = \delta_{i\alpha}$. We build a model analogous to
a set of 'trial' operators (linear
combinations of the ideal operators) by multiplying $\tilde{Z}$ by a
random orthogonal $10 \times 10$ matrix, $Z^\alpha_i
=\tilde{Z}^\alpha_j M^j_i$. We construct a non-diagonal correlator
matrix with elements
\begin{equation}
C_{ij}(t) = \sum_\alpha \frac{Z^{\alpha*}_i Z^\alpha_j}{2m_\alpha}
e^{-m_\alpha t}.
\end{equation}
 Application of the variational method to the full $10\times 10$
 matrix of correlators solves the problem exactly for any value of
 $t_0$. The situation we deal with in practice is not like this - we do
 not have anything like a 'complete' basis of operators, we can model
 this by truncating the correlator matrix down to a $5\times 5$
 submatrix.

With a selection of random orthogonal rotations ($M^j_i$) we observe a range of
behaviors in the solution to the variational problem, here we show two
such cases.

\subsection{Case 1}
\begin{figure}[h]
  \vspace{1cm}
       \psfig{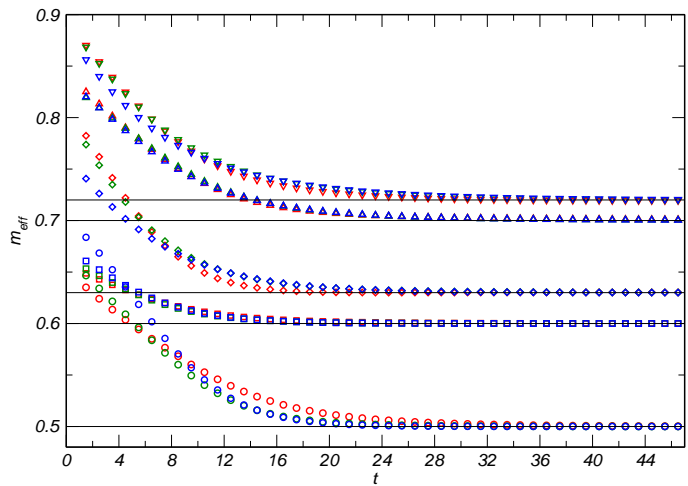}
      \caption{Extracted meff $\lambda(t)$ for $t_0=2$(red), $t_0=10$(green) and $t_0=30$(blue). Horizontal black lines are the input spectrum.  }
  \label{fig:toy1meff}
\end{figure}

\begin{figure}[h]
  \vspace{1cm}
       \psfig{width=15cm,file= 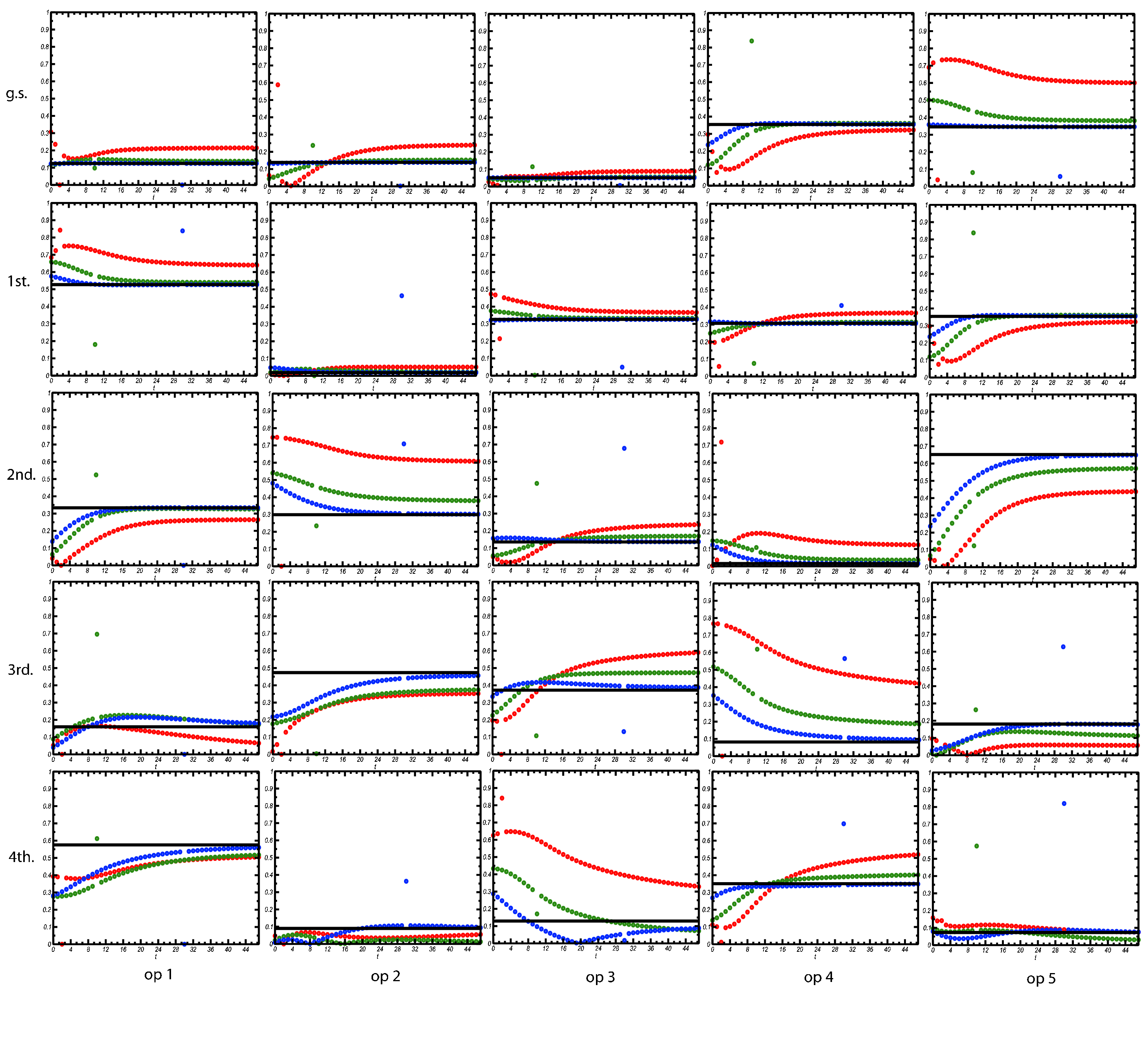}
      \caption{Extracted $|Z(t)|$ for $t_0=2$(red), $t_0=10$(green) and $t_0=30$(blue). Horizontal black line is the input Z.  }
  \label{fig:toy1Z}
\end{figure}

In figure \ref{fig:toy1meff} we show the effective masses of the principal
correlators for the choices $t_0 = 2, 10, 30$. We see that there is
relatively little sensitivity in this case to the value of $t_0$. On
the other hand we see considerable sensitivity to $t_0$ in the $Z(t)$
values shown in figure \ref{fig:toy1Z} and only approach accurate determination of
the $Z$ for the largest $t_0$ value.

\subsection{Case 2}
\begin{figure}[h]
  \vspace{1cm}
       \psfig{width=9cm,file= 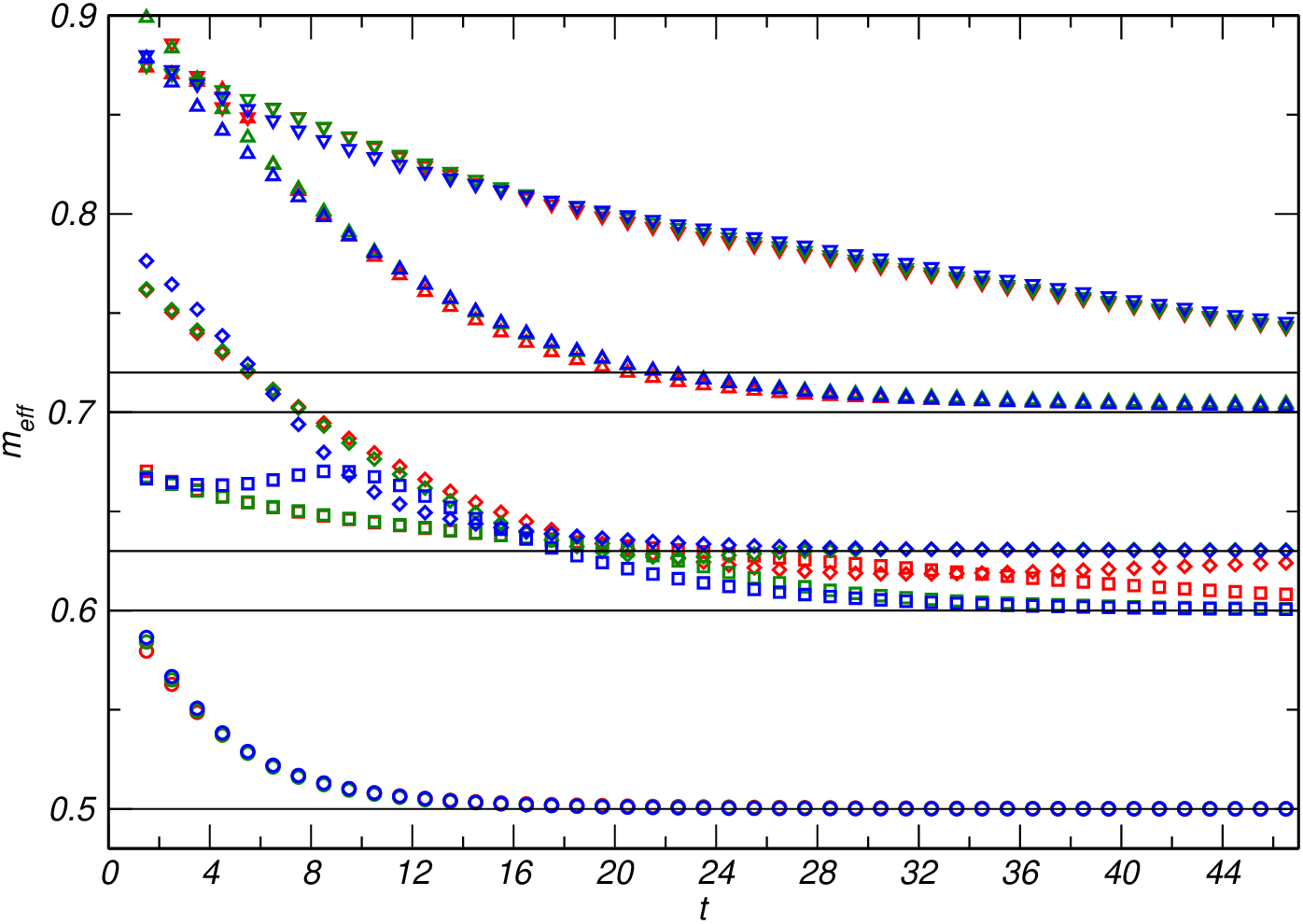}
      \caption{Extracted meff $\lambda(t)$ for $t_0=2$(red), $t_0=10$(green) and $t_0=30$(blue). Horizontal black lines are the input spectrum.  }
  \label{fig:toy2meff}
\end{figure}

In figure \ref{fig:toy2meff} we show the effective masses of the principal
correlators for the choices $t_0 = 2, 10, 30$, note the qualitative
differences with respect to the previous case including a flip in
level ordering and the lack of a plateau in the 4th excited state. The
plateaus of the first and second excited states clearly improve as
$t_0$ is increased. 

\bibliography{charm_spectrum} 

\end{document}